\documentclass{article}

    \PassOptionsToPackage{numbers, compress}{natbib}
\usepackage{wrapfig}

\usepackage[final]{neurips_2024}
\usepackage{amsmath} 
\usepackage{graphicx} 
\usepackage{booktabs,longtable,multirow,array}
\usepackage{amssymb} 

\usepackage{tikz}
\usetikzlibrary{arrows,positioning,calc,fit} % robust + widely available

\tikzset{
  >=stealth, % use classic stealth tip (works without arrows.meta)
  layer/.style   ={draw, rounded corners=3pt, semithick, fill=gray!6,
                   align=center, minimum width=48mm, minimum height=9mm, font=\small},
  hiedge/.style  ={->, line width=0.9pt},
  xedge/.style   ={<->, dashed, line width=1.2pt},
  blocked/.style ={->, dotted, line width=0.9pt},
  legendbox/.style={draw, rounded corners=3pt, fill=gray!6, inner sep=2mm, align=left, font=\small},
  insetbox/.style ={draw, rounded corners=3pt, fill=gray!5, inner sep=3mm}
}

\usepackage{caption}

\usepackage[utf8]{inputenc} % allow utf-8 input
\usepackage[T1]{fontenc}    % use 8-bit T1 fonts
\usepackage{hyperref}       % hyperlinks
\usepackage{url}            % simple URL typesetting
\usepackage{amsfonts}       % blackboard math symbols
\usepackage{nicefrac}       % compact symbols for 1/2, etc.
\usepackage{microtype}      % microtypography
\usepackage{xcolor}         % colors

\usepackage{etoolbox}  
\title{Agent Mars:\\ Multi-Agent Simulation for Multi-Planetary Life Exploration and Settlement}

\author{%
  Ziyang Wang\\
  {\tt\small ziyangwang@ieee.org}\\
}

\begin{document}

\maketitle

\noindent
\begin{minipage}{\textwidth}
  \centering
  \includegraphics[width=\textwidth]{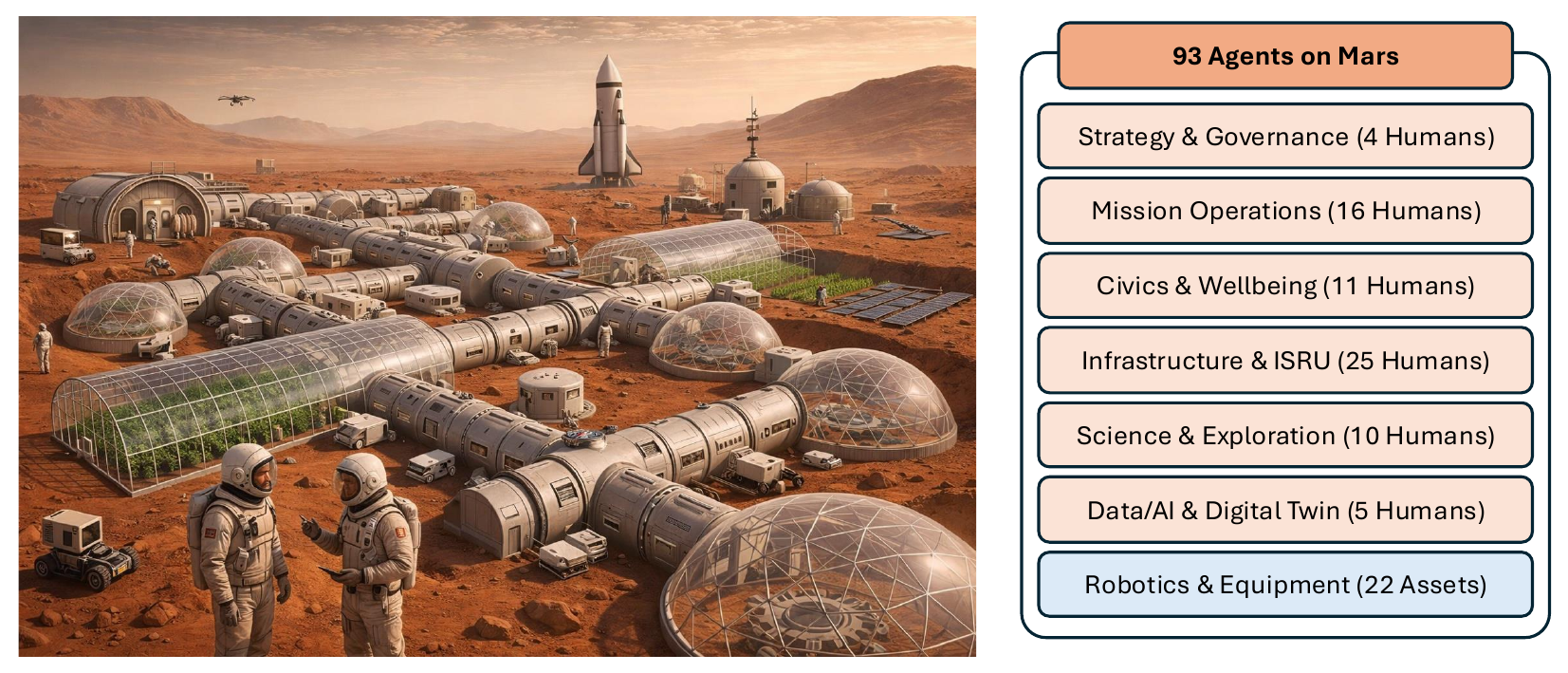}
  \captionof{figure}{Conceptual Mars base operations (left) and the Agent Mars roster of 93 agents organised into 7 layers of command and execution (right).}
  \label{fig:first}
\end{minipage}

\begin{abstract}
Artificial Intelligence (AI) has delivered transformative advances across agriculture, robotics, healthcare, industry, creative arts and science discovery yet its most consequential frontier may lie beyond Earth. Space exploration and settlement offer access to vast environments and resources that could fundamentally expand humanity’s economic and scientific horizon, but they also impose operational constraints unmatched by terrestrial settings: delayed and intermittent communications, extreme resource scarcity, heterogeneous expertise, and stringent requirements for safety, accountability and command authority. A central, unsolved challenge is therefore not isolated autonomy, but auditable coordination among many specialised humans, robots and digital services operating as a safety-critical system-of-systems. This paper introduces Agent Mars, an open, end-to-end multi-agent simulation framework for Mars base operations. Agent Mars begins by formalising a realistic operational organisation: a 93-agent roster spanning seven layers of command and execution (including human roles and physical assets), enabling controlled studies of coordination at base scale rather than toy multi-agent settings. Building on this foundation, Agent Mars implements a hierarchical \& cross-layer coordination paradigm that preserves chain-of-command while allowing explicitly vetted cross-layer exchanges with audit trails; supports dynamic role handover for resilient asset control with automatic failover under outages; and enables phase-dependent leadership that adapts command to daily operations, emergencies and science campaigns. Beyond organisational structure, Agent Mars models mission-critical interaction mechanisms—scenario-aware short- and long-horizon memory, configurable propose–vote consensus, and translator-mediated heterogeneous technical protocols—to capture how multidisciplinary teams align under stress. To quantify system-level behaviour, we propose the Agent Mars Performance Index (AMPI), an interpretable composite score that summarizes efficiency and robustness while retaining diagnostic component metrics. Across a released suite of 13 reproducible Mars-relevant operational scripts spanning life support, power, in-situ resource utilization, EVA and communications, logistics and science, Agent Mars reveals measurable coordination trade-offs and identifies regimes where curated cross-layer collaboration and functional leadership reduce coordination overhead without sacrificing reliability. Agent Mars establishes a benchmarkable, auditable foundation for Space AI, enabling systematic development of multi-agent autonomy for multi-planetary operations with implications for other high-stakes, resource-bounded domains.

\end{abstract}

\section{Introduction}
\label{sec:introduction}

AI has catalyzed transformative advances across agriculture, biology, manufacturing, medicine, and the creative arts. From precision farming and protein folding to industrial automation and generative art, AI-driven systems are delivering tangible benefits in efficiency and discovery. However, space exploration, a frontier with immense potential resources and critical societal importance, remains relatively under-explored by modern AI \cite{wang2025space}. The timing is opportune: the space economy and investment are growing rapidly, with renewed commercial momentum for lunar and planetary missions. Unlike terrestrial deployments, planetary operations must function under long communication delays to Earth ($\approx$ 4–21 minutes one-way), tight mass/energy margins, and safety-critical constraints. Space systems are inherently systems-of-systems encompassing humans, robots, software services, and physical infrastructure that must work in concert under autonomy. This complexity calls for advanced autonomous Multi-Agent Systems (MAS), namely heterogeneous teams of robotic and virtual agents operating with human oversight, that can coordinate operations, adapt to failures, and maintain a strict chain of command while conserving bandwidth. Recent breakthroughs in large language models (LLMs) make such autonomy increasingly feasible: LLM-based agents can reason in natural language, plan multi-step procedures, leverage world knowledge, and operate tools or simulations through text interfaces, enabling the development of AI crew members that can be tested in silico before deployment on real missions \cite{liu2024deepseek,liu2024deepseek2,team2023gemini,achiam2023gpt4,yang2025qwen3}.

A fast-growing body of research demonstrates that LLM-driven agents can plan, communicate, and collaborate in open-ended environments. Generative Agents integrate an LLM with long-term memory and reflective capabilities to produce believable, coordinated behaviors among dozens of simulated characters in a sandbox town \cite{park2023generative}. Domain-specific simulacra push this concept further: \emph{AI Hospital} benchmarks LLMs in multi-agent medical interaction settings \cite{fan2025ai}, while \emph{Agent Hospital} models a simulacrum of hospital workflows populated by evolvable medical agents \cite{li2024agenthospital}. \emph{Agent Laboratory} coordinates LLM researcher agents through literature review, experimentation, and report writing \cite{schmidgall2025agentlab}. Beyond individual environments, general multi-agent orchestration frameworks have emerged. \emph{AutoGen} \cite{wu2024autogen}, \emph{CAMEL} \cite{li2023camel}, \emph{MetaGPT} \cite{hong2023metagpt}, and \emph{ChatDev} \cite{qian2023chatdev} demonstrate that specialized LLM agents (e.g., Manager, Engineer, Tester) can decompose complex problems into sub-tasks, communicate via natural language, and jointly produce artifacts such as software repositories. Methodologically, techniques such as \emph{ReAct} \cite{yao2023react} and \emph{Reflexion} \cite{shinn2023reflexion} combine reasoning with acting and self-correction, while systems such as \emph{Voyager} \cite{wang2023voyager} augment agents with skill libraries and long-horizon memory to enable continual learning. These works reveal emergent collaboration, tool use, and problem-solving by LLM agents in a variety of settings. However, current LLM-agent frameworks largely operate in unconstrained sandbox environments: they often lack an explicit hierarchical chain-of-command, enforce few safety or resource constraints, and have not been evaluated under the mission-critical, resource-bounded conditions characteristic of planetary operations. In other words, existing studies illustrate what LLM-based agents can do, but not how to make them reliably do the right thing under strict operational rules and real mission failure modes.

Meanwhile, the MAS community offers decades of insight into reliable coordination under constraints \cite{wooldridge2009introduction}. Classical multi-robot coordination developed formal approaches for task allocation, planning, and resilience in distributed teams. The well-known multi-robot task allocation (MRTA) taxonomy defines which agents should do what tasks, when, and where, providing a foundation for optimal and heuristic task assignment \cite{gerkey2004formal}. Distributed constraint optimization and planning algorithms (e.g., ADOPT \cite{modi2005adopt}, DPOP \cite{petcu2005scalable}) and Dec-POMDP formulations \cite{oliehoek2016concise} tackle decision-making under uncertainty and intermittent communication. Multi-agent reinforcement learning (MARL) has yielded scalable policies and learned communication protocols---for example, centralized training with decentralized execution in MADDPG \cite{lowe2017multi}, value factorization in QMIX \cite{rashid2020monotonic}, policy-gradient methods such as MAPPO \cite{yu2022surprising}, and differentiable communication models including CommNet \cite{sukhbaatar2016learning}, DIAL \cite{foerster2016learning}, and TarMAC \cite{das2019tarmac}. Fielded systems further demonstrate robustness in harsh domains: notably, teams in the DARPA Subterranean Challenge coordinated heterogeneous robots to autonomously explore extreme environments under severe communication blackouts \cite{chung2023into,tranzatto2022cerberus,agha2021nebula}. These classical frameworks show that multi-agent autonomy can be safe and effective in unpredictable conditions. However, they typically rely on fixed interfaces or narrow learned policies and lack the flexible, high-level reasoning and cross-domain adaptability that language-based intelligence provides. In essence, a gap remains between the reliability of traditional multi-agent control and the generality of LLM-based reasoning.

Spaceflight autonomy technologies to date reflect this trade-off. Timeline-based planning and robust execution systems are standard: NASA planning frameworks such as ASPEN \cite{fukunaga1997aspen} and EUROPA \cite{barreiro2012europa}, and execution languages such as PLEXIL \cite{verma2005plan} provide deterministic, verifiable cores for on-board plan execution. Modern flight software frameworks (e.g., NASA's Core Flight System, cFS \cite{mccomas2016core}, and JPL's \texttt{F$^\prime$} flight software \cite{bocchino2018f}) emphasize fault tolerance, predictable behavior, and integration with low-level hardware---all essential for safety-critical operations. These systems excel at enforcing constraints, scheduling activities under resource limits, and reacting to events in controlled ways. Yet, they operate far from the abstract mission reasoning level and offer limited interfaces for natural language or ``common sense'' knowledge. Integrating high-level reasoning (e.g., diagnosing a cascading anomaly by synthesizing evidence across geology, life support, and power) with low-level reliability remains an open challenge. Bridging this gap is crucial for future planetary bases, where autonomous teams will require both robust low-level control and flexible high-level decision-making across diverse disciplines.

In this work, we present Agent Mars, a Mars-base multi-agent simulation framework for studying settlement-scale operations. Specifically, we contribute:

\begin{enumerate}
\item a 93-agent roster spanning seven operational layers with explicit role/asset ownership;
\item an auditable Hierarchical \& Cross-Layer Coordination (HCLC) architecture with a strict chain-of-command by default, whitelisted cross-layer shortcuts, and dynamic role handover;
\item interaction modules for scenario-aware memory, configurable propose–vote consensus, and translator-mediated cross-specialty communication;
\item a 13-scenario benchmark suite with fixed prompts and an interpretable system-level metric, Agent Mars Performance Index (AMPI), for comparing organizational strategies.
\end{enumerate}

\begin{figure}[htbp]
    \centering
    \includegraphics[width=\linewidth]{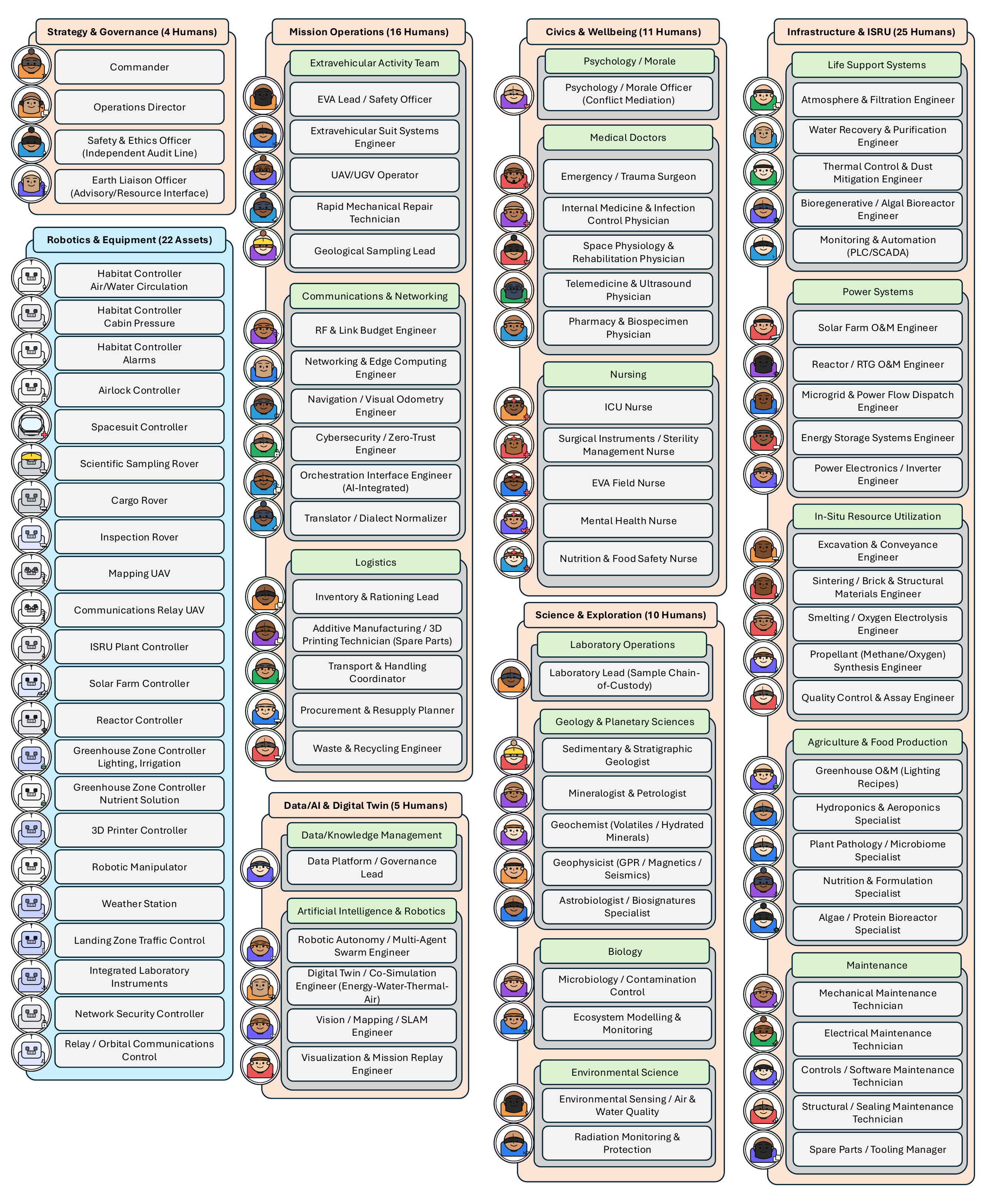}
    \caption{Roster on Mars. 93 agents organized into 7 layers with 18 functional groups.}
    \label{fig:roster}
\end{figure}

\section{Agent Mars}
\label{sec:agent-mars}

Agent Mars is an open, controllable multi-agent simulation testbed for settlement-scale Mars base operations with 93 agents (see Fig. \ref{fig:first}). A run is driven by a scenario script and executed by a roster of heterogeneous agents (human role agents and asset controllers) organised into a seven-layer chain of command; agents interact through a message router that enforces hierarchical routing by default and logs any audited cross-layer exchanges. This abstraction captures mission-critical coordination constraints (authority, accountability, and communication bottlenecks) while enabling systematic, repeatable comparisons of alternative coordination policies.

This section presents Agent Mars in four parts: (i) the curated roster and explicit role/asset ownership (\S\ref{sec:roster-mars}); (ii) the hierarchical \& cross-layer coordination mechanism (HCLC) for routing, handover, and leadership (\S\ref{subsec:hclc}); (iii) interaction modules for memory, consensus, and translator-mediated protocols (\S\ref{subsec:interaction}); and (iv) the benchmark task suite and AMPI metric used for evaluation (\S\ref{subsec:bench-ampi}).

\subsection{Roster of Agents}
\label{sec:roster-mars}
\begin{wrapfigure}[17]{r}{0.28\textwidth}
  \centering
  \includegraphics[width=\linewidth]{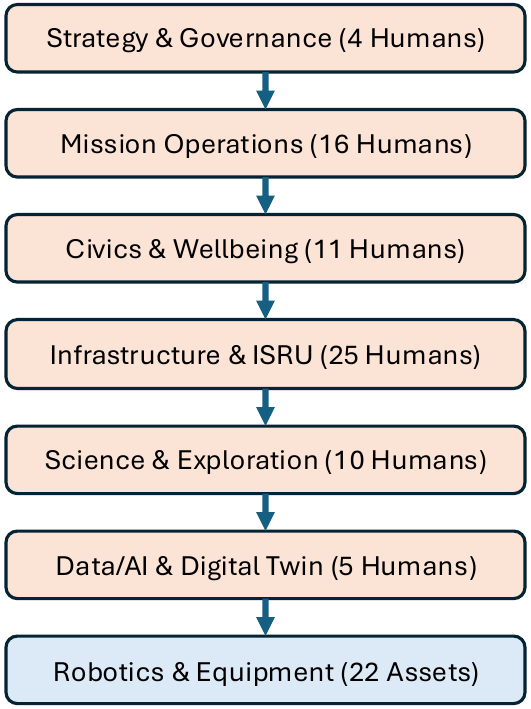}
  \caption{7-layer hierarchy.}
  \label{fig:Hierarchy}
\end{wrapfigure}

As government-led lunar programmes and commercial spaceflight roadmaps move from single missions to sustained surface operations, the centre of gravity is shifting from vehicle autonomy to settlement-scale operations: continuous life support, power generation and storage, in-situ resource utilisation, logistics, EVA, science, and cyber-physical security. These activities must be coordinated under delayed and intermittent communications, strict safety authority, and tight resource margins, making end-to-end validation on hardware infeasible and turning simulation into a primary instrument for system design and risk reduction. Agent Mars therefore instantiates a base-scale operational organisation with 93 agents (71 human role agents and 22 asset controllers; Fig.~\ref{fig:roster}). The roster is structured as a 7-layer hierarchy (Fig. \ref{fig:Hierarchy}) with 18 functional groups, reflecting a pyramid of accountability in which \emph{Strategy \& Governance} sets intent and authorises actions, \emph{Mission Operations} translates intent into executable plans, downstream functional layers execute and monitor, and the \emph{Assets} layer interfaces with physical systems.  The complete role specification of human and assets is provided in Table~\ref{tab:abbr-human} and Table \ref{tab:abbr-assets}, respectively. To make responsibilities explicit, each asset is assigned a named primary operator and a designated backup illustrated in Table~\ref{tab:ownership-full}, mirroring operational handover practice and ensuring that control authority is always defined.

\noindent\textbf{Strategy \& Governance} – This top layer establishes the mission’s intent, risk posture, and authority structure. It comprises the Base Commander (overall mission lead), an Operations Director (managing day-to-day execution), a Safety \& Ethics Officer (independent oversight for crew safety and ethical compliance), and an Earth Liaison (interface for Earth-based support, commercial resupply, and advisory input). Together, these roles set high-level strategy, approve plans, and provide a clear command hierarchy analogous to those in real lunar or Martian base proposals.

\textbf{Mission Operations} – The operations layer coordinates daily activities across extravehicular, communications, and logistics functions. The EVA team handles all extravehicular activity, including an EVA lead (safety officer), a spacesuit systems specialist, an unmanned vehicle (UxV) operator for drones/rovers, a rapid repair technician for field fixes, and a sampling lead for collecting geological specimens. The communications group maintains all networks and navigation support: it covers radio frequency link budgeting and relay scheduling, on-site networking and edge computing, navigation and visual odometry for localization, cybersecurity under a zero-trust model, and AI-driven scheduling/dispatch interfaces. The logistics group manages resources and supply lines: this includes inventory control and rationing, additive manufacturing (3D printing) of spare parts, waste management and recycling for closed-loop sustainability, material transport and handling, and procurement planning for resupply missions. Mission Operations thus ensures that plans from leadership translate into coordinated action in the habitat and field.

\textbf{Civics \& Wellbeing} – This layer safeguards crew health, welfare, and cohesion. It includes a multi-disciplinary medical staff with physicians specialized in trauma and emergency care, internal medicine and infection control, space physiology and rehabilitation, telemedicine (e.g. ultrasound diagnostics) and a pharmacist/biobank manager – collectively capable of handling medical emergencies and long-term health monitoring. Supporting them is a nursing staff composed of ICU nurses, sterile instrument technicians, EVA field medics, mental health nurses, and nutritionists to oversee food safety and diet. A psychology officer focuses on crew morale, mental health, and conflict mediation. By encompassing physical health, mental well-being, and social dynamics, this layer addresses the human factors crucial for a successful long-duration mission.

{\bf Infrastructure \& ISRU} – This layer maintains the base’s critical systems and local resource utilization. The life-support group manages habitat environmental control, including atmospheric composition and filtration, water recovery and purification, thermal regulation and dust mitigation, bioregenerative systems (e.g. algal oxygenators or greenhouses), and automated monitoring via industrial control systems (PLC/SCADA). The power group oversees energy generation and distribution, from solar farm operations and dust cleaning, to nuclear reactor or RTG management, microgrid control and load scheduling, battery energy storage, and power electronics/inverter maintenance. The ISRU group (In-Situ Resource Utilization) handles extraction and production of local resources: mining and regolith conveyance, regolith sintering and construction material fabrication, metal smelting and oxygen electrolysis, propellant synthesis (e.g. methane/oxygen for rockets), and quality assurance of outputs. Agriculture specialists tend to food production and related biocycles, optimizing greenhouse light spectra, hydroponic/aeroponic cultivation, plant pathology prevention and microbiome control, crop nutrition and processing, and algae/protein bioreactor systems for food and oxygen. Maintenance technicians provide mechanical, electrical, and software upkeep of equipment, ensure structural integrity (e.g. habitat seals), and manage spare parts and tooling. Together, these five sub-groups keep the habitat functional and increasingly self-sustaining by producing vital resources on-site and fixing issues before they escalate.

{\bf Science \& Exploration} – Fulfilling the mission’s research objectives, this layer comprises teams for scientific investigation and exploration. The geology group includes experts in sedimentology/stratigraphy (to read the planet’s geological history), mineralogy and petrology (to characterize rocks and minerals), geochemistry of volatiles and hydrated minerals (to analyze resource-bearing compounds and signs of water alteration), geophysics (for ground-penetrating radar, magnetics, seismic surveying of the subsurface), and astrobiology (to search for biosignatures or past life). The biology group focuses on life sciences, including a microbiologist concerned with contamination control and studying microbial life, and an ecologist to model and monitor the closed ecosystem of the habitat. The environmental science role monitors habitat and ambient conditions, tracking air and water quality, radiation levels, and ensuring radiation protection measures. A laboratory operations lead coordinates sample handling, curation, and analysis protocols to maintain the chain-of-custody and integrity of scientific samples. This science layer ensures that exploration and research – from field sampling to lab analysis – are carried out rigorously, contributing to the broader goals of the mission.

{\bf Data/AI \& Digital Twin} – This layer provides advanced computational support, autonomy, and modeling capabilities for the mission. It includes a data platform and governance lead who oversees data infrastructure and policies, ensuring that information flows securely and efficiently. A robotics autonomy engineer develops and supervises the AI for robotic agents and multi-robot swarms (such as coordinating rovers or drones in the field). There are specialists maintaining digital twins – high-fidelity simulations of the base’s coupled systems (energy, water, thermal, air) used to predict system behavior and optimize operations. A computer vision and mapping expert (SLAM – Simultaneous Localization and Mapping) handles autonomous navigation and environmental mapping, and a visualization/mission replay engineer produces interfaces for situational awareness, replaying events or simulations for debrief and training. By integrating these AI and modeling roles, the Data/AI layer enhances decision-making, predictive maintenance, and autonomy, serving as a technological force-multiplier for the crew.

{\bf Robotic \& Equipment Assets} – The final layer consists of all the physical systems and robots represented as agents. This includes habitat environmental control units, airlocks, surface rovers (science rovers, inspection rovers, and cargo haulers), uncrewed aerial vehicles (for mapping and communication relay), the ISRU processing plant, power generation units (solar arrays and reactor control systems), greenhouse control systems, 3D printers and fabrication units, robotic manipulators, the weather monitoring station, landing zone traffic control systems, laboratory instruments, network security appliances, and a communications satellite relay. Each of these asset agents operates under the supervision of a designated human owner (from the layers above) with a backup controller also assigned. This one-to-one mapping of critical hardware to responsible humans reflects real mission protocols and ensures accountability and redundancy. By explicitly including assets as agents, Agent Mars captures the interactions between crew members and the equipment they operate, allowing the simulation to model how commands propagate to hardware and how autonomous systems report status or anomalies back to the crew. This holistic roster from Command down to individual machines enables us to explore governance, teamwork, and failure recovery in a realistic Mars base setting.

\begin{longtable}{@{}p{0.26\linewidth}p{0.70\linewidth}@{}}
\caption{Human role agents (N=71): IDs (abbrev.) and full role titles.}
\label{tab:abbr-human}\\
\toprule
\textbf{ID (abbrev.)} & \textbf{Full role title} \\
\midrule
\endfirsthead
\toprule
\textbf{ID (abbrev.)} & \textbf{Full role title} \\
\midrule
\endhead
\bottomrule
\endfoot

\addlinespace[0.6ex]
\multicolumn{2}{@{}l}{\textbf{Strategy \& Governance}}\\
\addlinespace[0.2ex]
\texttt{CMD\_01}   & Commander \\
\texttt{EARTH\_01} & Earth Liaison Officer (Advisory/Resource Interface) \\
\texttt{OPS\_01}   & Operations Director \\
\texttt{SEO\_01}   & Safety \& Ethics Officer (Independent Audit Line) \\

\addlinespace[0.8ex]
\multicolumn{2}{@{}l}{\textbf{Mission Operations}}\\
\addlinespace[0.2ex]
\texttt{COM\_04} & Cybersecurity / Zero-Trust Engineer \\
\texttt{COM\_03} & Navigation / Visual Odometry Engineer \\
\texttt{COM\_02} & Network \& Edge Computing Engineer \\
\texttt{COM\_01} & RF \& Link Budget Engineer \\
\texttt{COM\_05} & Scheduling Interface Engineer (AI-Integrated) \\
\texttt{COM\_06} & Translator / Dialect Normalizer \\

\texttt{EVA\_01} & EVA Lead / Safety Officer \\
\texttt{EVA\_04} & Rapid Mechanical Repair Technician \\
\texttt{EVA\_02} & Spacesuit Systems Engineer \\
\texttt{EVA\_03} & UAV/UGV Operator \\
\texttt{EVA\_05} & Geological Sampling Lead \\

\texttt{LOGT\_02} & AM/3D Printing Technician (Spares) \\
\texttt{LOGT\_05} & Procurement \& Resupply Planner \\
\texttt{LOGT\_04} & Transport \& Handling Coordinator \\
\texttt{LOGT\_01} & Inventory \& Rationing Lead \\
\texttt{LOGT\_03} & Waste \& Recycling Engineer \\

\addlinespace[0.8ex]
\multicolumn{2}{@{}l}{\textbf{Civics \& Wellbeing}}\\
\addlinespace[0.2ex]
\texttt{MED\_01} & Emergency/Trauma Surgeon \\
\texttt{MED\_02} & Internal Medicine \& Infection Control Physician \\
\texttt{MED\_05} & Pharmacology \& Biobanking Physician \\
\texttt{MED\_03} & Space Physiology \& Rehabilitation Physician \\
\texttt{MED\_04} & Telemedicine \& Ultrasound Physician \\

\texttt{NUR\_03} & EVA Field Nursing \\
\texttt{NUR\_01} & ICU Nursing \\
\texttt{NUR\_04} & Mental Health Nursing \\
\texttt{NUR\_05} & Nutrition \& Food Safety \\
\texttt{NUR\_02} & Surgical Instruments / Sterile Processing \\

\texttt{PSY\_01} & Psychology/Morale Officer (Conflict Mediation) \\

\addlinespace[0.8ex]
\multicolumn{2}{@{}l}{\textbf{Infrastructure \& ISRU}}\\
\addlinespace[0.2ex]
\texttt{AGRI\_05} & Algae/Protein Bioreactor Specialist \\
\texttt{AGRI\_01} & Greenhouse O\&M (Light Recipes) \\
\texttt{AGRI\_02} & Hydroponics \& Aeroponics Specialist \\
\texttt{AGRI\_04} & Nutrition \& Formulation Specialist \\
\texttt{AGRI\_03} & Plant Pathology / Microbiome Specialist \\

\texttt{ISRU\_01} & Excavation \& Conveyance Engineer \\
\texttt{ISRU\_04} & Propellant (Methane/Oxygen) Synthesis Engineer \\
\texttt{ISRU\_05} & Quality Inspection \& Assay Engineer \\
\texttt{ISRU\_02} & Sintering/Brick \& Structural Materials Engineer \\
\texttt{ISRU\_03} & Smelting / Oxygen Electrolysis Engineer \\

\texttt{LSS\_01} & Atmosphere \& Filtration Engineer \\
\texttt{LSS\_04} & Bioregenerative / Algal Bioreactor Engineer \\
\texttt{LSS\_05} & Monitoring \& Automation (PLC/SCADA) \\
\texttt{LSS\_03} & Thermal Control \& Dust Mitigation Engineer \\
\texttt{LSS\_02} & Water Recycling \& Purification Engineer \\

\texttt{MNT\_03} & Controls/Software Maintenance Technician \\
\texttt{MNT\_02} & Electrical Maintenance Technician \\
\texttt{MNT\_01} & Mechanical Maintenance Technician \\
\texttt{MNT\_05} & Spares/Tooling Administrator \\
\texttt{MNT\_04} & Structural/Sealing Maintenance Technician \\

\texttt{PWR\_04} & Energy Storage Systems Engineer \\
\texttt{PWR\_03} & Microgrid \& Power Flow Dispatch Engineer \\
\texttt{PWR\_05} & Power Electronics / Inverter Engineer \\
\texttt{PWR\_02} & Reactor/RTG O\&M Engineer \\
\texttt{PWR\_01} & Solar Farm O\&M Engineer \\

\addlinespace[0.8ex]
\multicolumn{2}{@{}l}{\textbf{Science \& Exploration}}\\
\addlinespace[0.2ex]
\texttt{BIO\_02} & Ecosystem Modeling \& Monitoring \\
\texttt{BIO\_01} & Microbiologist / Contamination Control \\
\texttt{ENV\_01} & Environmental Sensing / Air \& Water Quality \\
\texttt{ENV\_02} & Radiation Monitoring \& Protection \\
\texttt{GEO\_05} & Astrobiologist / Biosignatures \\
\texttt{GEO\_03} & Geochemist (Volatiles/Hydrated Minerals) \\
\texttt{GEO\_04} & Geophysicist (GPR/Magnetics/Seismics) \\
\texttt{GEO\_02} & Mineralogist \& Petrologist \\
\texttt{GEO\_01} & Sedimentology \& Stratigraphy Geologist \\
\texttt{LAB\_01} & Laboratory Manager (Sample Chain) \\

\addlinespace[0.8ex]
\multicolumn{2}{@{}l}{\textbf{Data/AI \& Digital Twin}}\\
\addlinespace[0.2ex]
\texttt{AI\_03}  & Digital Twin / Co-Simulation Engineer (Energy--Water--Thermal--Air) \\
\texttt{AI\_02}  & Robotic Autonomy / Multi-Agent Swarm Engineer \\
\texttt{AI\_04}  & Vision/Mapping/SLAM Engineer \\
\texttt{AI\_05}  & Visualization \& Mission Replay Engineer \\
\texttt{DKM\_01} & Data Platform / Governance Lead \\

\end{longtable}

\begin{longtable}{@{}p{0.30\linewidth}p{0.66\linewidth}@{}}
\caption{Asset controller agents (N=22): IDs (abbrev.) and asset names.}
\label{tab:abbr-assets}\\
\toprule
\textbf{ID (abbrev.)} & \textbf{Asset name} \\
\midrule
\endfirsthead
\toprule
\textbf{ID (abbrev.)} & \textbf{Asset name} \\
\midrule
\endhead
\bottomrule
\endfoot

\texttt{AIRLOCK\_CTRL\_01}     & Airlock Controller \\
\texttt{ARM\_CTRL\_01}         & Robotic Manipulator \\
\texttt{ATC\_LZ\_01}           & Landing Zone Air/Traffic Control \\
\texttt{COMSAT\_CTRL\_01}      & Relay/Orbital Communications Controller \\
\texttt{GH\_CTRL\_01}          & Greenhouse Zone Controller (Lighting, Irrigation) \\
\texttt{GH\_CTRL\_02}          & Greenhouse Zone Controller (Nutrient Solution) \\
\texttt{HAB\_01}               & Habitat Controller (Air/Water Circulation) \\
\texttt{HAB\_02}               & Habitat Controller (Cabin Pressure) \\
\texttt{HAB\_03}               & Habitat Controller (Alarms) \\
\texttt{ISRU\_PLANT\_01}       & ISRU Plant Controller \\
\texttt{LAB\_INSTR\_01}        & Laboratory Instrument Suite \\
\texttt{NET\_SEC\_CTRL\_01}    & Network Security Controller \\
\texttt{NUKE\_CTRL\_01}        & Reactor Controller \\
\texttt{PRT\_CTRL\_01}         & 3D Printer Controller \\
\texttt{ROV\_CARGO\_01}        & Cargo Rover \\
\texttt{ROV\_INSP\_01}         & Inspection Rover \\
\texttt{ROV\_SCI\_01}          & Science Sampling Rover \\
\texttt{SOL\_CTRL\_01}         & Solar Farm Controller \\
\texttt{SUIT\_CTRL\_01}        & Spacesuit Controller \\
\texttt{UAV\_COM\_01}          & Communications Relay UAV \\
\texttt{UAV\_MAP\_01}          & Mapping UAV \\
\texttt{WX\_STATION\_01}       & Weather Station \\
\end{longtable}

Table~\ref{tab:ownership-full} lists the initialization-time mapping from human controllers to Layer-Assets, preserving functional groups and alarm levels from the roster. Assignments may change at runtime through incident response and role switching.

\begin{table}[t]
\centering
\tiny
\caption{Default mapping from human controllers to physical assets at initialization. Group and alarm level follow the current roster. Ownership may change at runtime via incident response and role handover.}
\label{tab:ownership-full}
\begin{tabular}{lllllp{0.38\linewidth}}
\toprule
\textbf{Asset (Layer: Assets)} & \textbf{Group} & \textbf{Primary (human)} & \textbf{Backup(s)} & \textbf{Alarm} & \textbf{Function (brief)} \\
\hline
HAB\_01 & HAB & LSS\_05 & LSS\_01 & A & Air/water circulation; cabin pressure; alarms \\
HAB\_02 & HAB & LSS\_05 & LSS\_02 & A & Air/water circulation; cabin pressure; alarms \\
HAB\_03 & HAB & LSS\_05 & LSS\_03 & A & Air/water circulation; cabin pressure; alarms \\
AIRLOCK\_CTRL\_01 & AIRLOCK & EVA\_01 & MNT\_04 & A & Interlocks; depress/repress; contamination control \\
SUIT\_CTRL\_01 & SUIT & EVA\_02 & NUR\_03 & A & PLSS telemetry; consumables/time remaining; leak alarms \\
ROV\_SCI\_01 & ROV\_SCI & GEO\_05 & EVA\_05 & B & Drilling; sampling cache; imaging \\
ROV\_CARGO\_01 & ROV\_CARGO & LOGT\_04 & MNT\_01 & B & Payload transport; follow navigation \\
ROV\_INSP\_01 & ROV\_INSP & MNT\_03 & MNT\_02 & B & Structural inspection; IR imaging \\
UAV\_MAP\_01 & UAV\_MAP & COM\_03 & AI\_04 & C & Aerial mapping; photogrammetry \\
UAV\_COM\_01 & UAV\_COM & COM\_02 & COM\_01 & B & Relay; ad-hoc network \\
ISRU\_PLANT\_01 & ISRU\_PLANT & ISRU\_03 & ISRU\_04 & A & Mining; electrolysis; methane synthesis \\
SOL\_CTRL\_01 & SOL\_CTRL & PWR\_01 & PWR\_05 & B & Sun tracking; panel cleaning; inverter control \\
NUKE\_CTRL\_01 & NUKE\_CTRL & PWR\_02 & PWR\_03 & A & Power dispatch; safe shutdown (SCRAM) \\
GH\_CTRL\_01 & GH\_CTRL & AGRI\_01 & AGRI\_02 & B & Lighting; irrigation; nutrient dosing \\
GH\_CTRL\_02 & GH\_CTRL & AGRI\_03 & AGRI\_04 & B & Lighting; irrigation; nutrient dosing \\
PRT\_CTRL\_01 & PRT\_CTRL & LOGT\_02 & MNT\_03 & C & Additive manufacturing queue; material/nozzle change \\
ARM\_CTRL\_01 & ARM\_CTRL & MNT\_01 & MNT\_02 & B & Assembly; maintenance; material handling \\
WX\_STATION\_01 & WX & ENV\_02 & COM\_01 & B & Storm alerts; dust concentration; irradiance \\
ATC\_LZ\_01 & ATC & OPS\_01 & COM\_05 & B & Landing-zone traffic control; windows; no-fly zones \\
LAB\_INSTR\_01 & LAB\_INSTR & LAB\_01 & MED\_05 & C & GC/MS; XRF; cold-chain management \\
NET\_SEC\_CTRL\_01 & NETSEC & COM\_04 & DKM\_01 & A & Whitelisting; segmentation; key rotation \\
COMSAT\_CTRL\_01 & COMSAT & COM\_01 & COM\_02 & A & Bandwidth allocation; link switching \\
\bottomrule
\end{tabular}
\end{table}

\subsection{Hierarchical \& Cross-Layer Coordination (HCLC)}
\label{subsec:hclc}

Coordinating large, heterogeneous agent teams involves a fundamental trade-off between hierarchical control and decentralized communication. Classical mission and safety-critical systems favor strict chains of command to preserve accountability and risk containment, while fully peer-to-peer coordination improves responsiveness at the cost of discipline. This tension is amplified in Mars base operations, where delayed and intermittent communications, safety authority, and tightly coupled infrastructure require both rapid cross-disciplinary exchange and explicit command responsibility. 

Agent Mars addresses this challenge through HCLC, a coordination paradigm tailored to settlement-scale planetary operations. HCLC preserves a formal chain of command as the default communication backbone, while allowing explicitly whitelisted and auditable cross-layer exchanges for time-critical interactions. Beyond routing, HCLC incorporates dynamic role switching for resilient asset control and mission-phase-aware leadership selection, enabling command to adapt across routine operations, emergencies, and science campaigns without compromising accountability. The mechanisms described below are implemented in the runtime router, leader selector, role-handoff controller, and metrics/AMPI modules, and are evaluated systematically in the Mars benchmark scenarios.

\paragraph{Cross-layer routing under a hierarchy.}
Strict hierarchies reduce risk but can inflate latency and message load; unconstrained peer-to-peer
links reduce delay but erode safety gates. We seek a middle ground that preserves hierarchical
accountability while enabling explicitly whitelisted cross-layer exchanges for time-critical tasks
(e.g., GEO$\rightarrow$AI for path planning).

Let $\mathcal{G}=(\mathcal{V},\mathcal{E}_H)$ be the directed hierarchical communication graph, where
$\mathcal{V}$ is the set of all agents (human roles and asset controllers) and $\mathcal{E}_H$ is the
set of directed edges representing the strict chain-of-command (e.g., Command $\rightarrow$
Operations $\rightarrow$ Functional teams $\rightarrow$ Assets). Let $g:\mathcal{V}\rightarrow
\mathrm{Groups}$ map each agent to its functional group (e.g., GEO, AI, COM).

We specify a group-level whitelist $W\subseteq \mathrm{Groups}\times\mathrm{Groups}$ of allowed
cross-layer shortcuts, such as (GEO,AI) or (GEO,COM) (Table~\ref{tab:whitelist}). The induced set of
whitelisted agent-to-agent edges is
\[
\mathcal{E}_X(W)=\{(u,v)\in\mathcal{V}\times\mathcal{V}:\ (g(u),g(v))\in W,\ g(u)\neq g(v)\}.
\]

\noindent\textbf{Routing rule.}
Under \texttt{STRICT}, the router allows only hierarchical edges $\mathcal{E}_H$.
Under \texttt{CROSSLAYER}, the router allows hierarchical edges plus whitelisted shortcuts
$\mathcal{E}_H\cup \mathcal{E}_X(W)$.
If a direct hop $(u,v)$ is \emph{not allowed} by the current policy (i.e., it is not in the allowed
edge set), the router forwards via an intermediate hub $h\in\{\mathrm{OPS},\mathrm{CMD}\}$ (Mission
Operations or Command), producing an auditable two-hop path $u\rightarrow h\rightarrow v$.
Formally, letting $\mathcal{E}_{\mathrm{allow}}=\mathcal{E}_H$ for \texttt{STRICT} and
$\mathcal{E}_{\mathrm{allow}}=\mathcal{E}_H\cup\mathcal{E}_X(W)$ for \texttt{CROSSLAYER}, the router
path function is
\[
\pi(u\rightarrow v)=
\begin{cases}
(u,v), & \text{if } (u,v)\in \mathcal{E}_{\mathrm{allow}},\\
(u,h),(h,v), & \text{otherwise},\ \ h\in\{\mathrm{OPS},\mathrm{CMD}\}.
\end{cases}
\]

Let $N_{\mathrm{msg}}$ be the total number of messages exchanged and $N_{\mathrm{cross}}$ the number
of messages that traverse an edge in $\mathcal{E}_X(W)$ under \texttt{CROSSLAYER}. The cross-layer
utilization ratio is
\[
\rho_{\mathrm{cross}}=\frac{N_{\mathrm{cross}}}{N_{\mathrm{msg}}}\in[0,1].
\]
At runtime, policy switching, whitelist loading, hub-forwarding, and message accounting are enforced
by the router and message loop; all hub-forwarded and cross-layer events are logged for audit.

Let $N_{\mathrm{msg}}$ be the total number of messages exchanged and $N_{\mathrm{cross}}$ the number of messages traversing edges in $\mathcal{E}_X(W)$.  
The cross-layer utilization ratio is defined as
\[
\rho_{\mathrm{cross}} = \frac{N_{\mathrm{cross}}}{N_{\mathrm{msg}}} \in [0,1],
\]
which can optionally be included in AMPI to penalize overuse of cross-layer shortcuts when desired.

\begin{table}[t]
\centering
\small
\caption{Default cross-layer shortcut whitelist $W$ at the group level. Direction is Source$\to$Target; all shortcuts are audited by the router.}
\label{tab:whitelist}
\begin{tabular}{@{}ll@{}}
\toprule
Source group & Allowed direct targets \\
\midrule
GEO  & AI, COM, LAB \\
BIO  & AI \\
LAB  & AI \\
COM  & AI \\
LSS  & AI \\
PWR  & AI \\
ISRU & AI \\
AGRI & AI \\
MNT  & AI \\
\bottomrule
\end{tabular}
\end{table}

\begin{figure}[htbp]
    \centering
    \includegraphics[width=\linewidth]{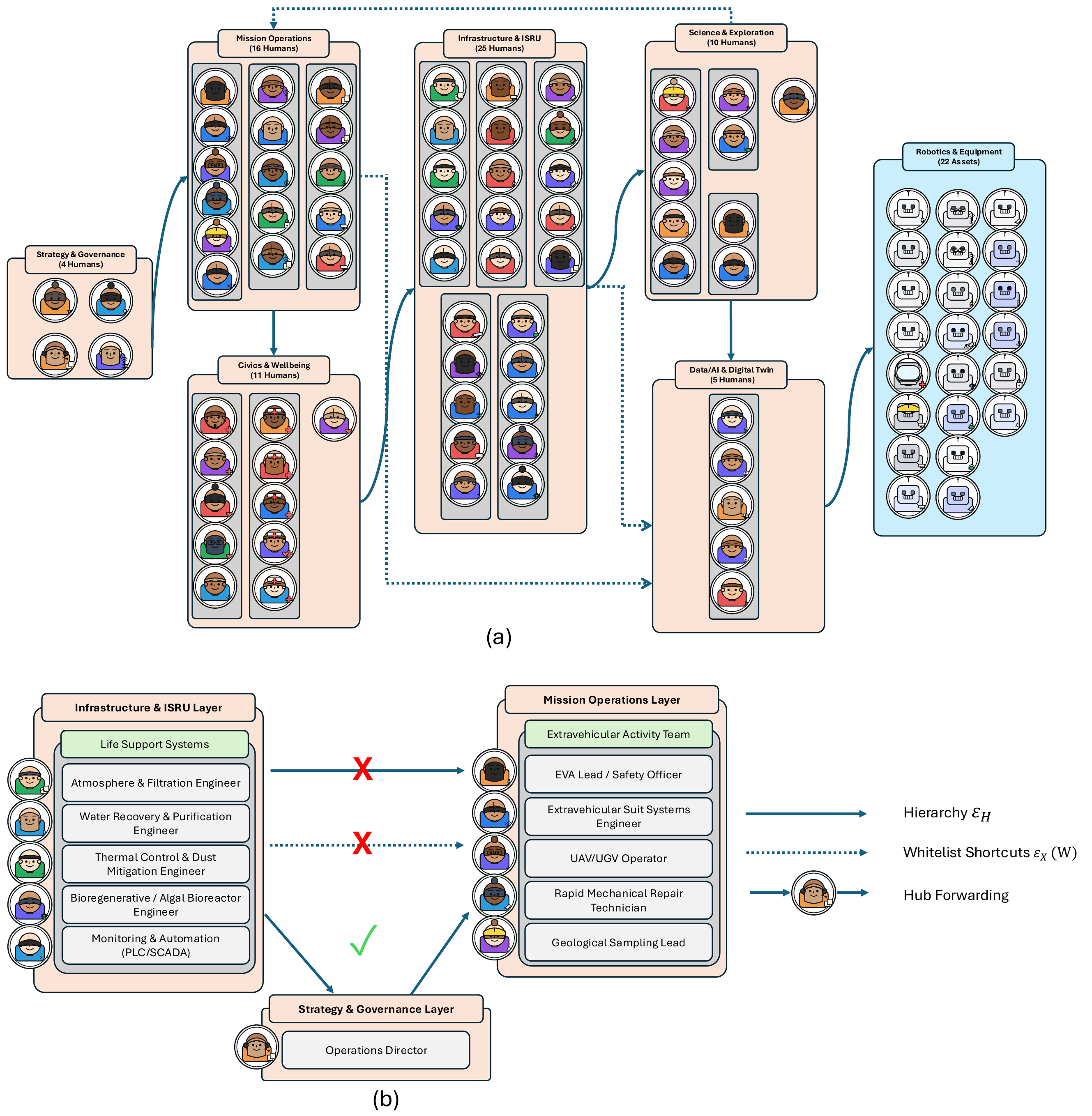}
    \caption{HCLC routing with curated shortcuts and audited hub-forwarding. 
    (a) Solid edges ($\mathcal{E}_H$) implement the strict hierarchy. Dashed arcs are representative whitelist shortcuts $\mathcal{E}_X(\mathcal{W})$ consistent with the default code configuration (e.g., GEO$\rightarrow$COM, COM$\rightarrow$AI, and LSS/PWR/ISRU/AGRI/MNT$\rightarrow$AI).
    (b) When $(u,v)\notin\mathcal{W}$, traffic is forwarded via the mission hub (OPS) with audit logging.}
    \label{fig:HCLC1}
\end{figure}

\paragraph{Redundancy and dynamic role switching for assets.}
Asset owners can be unavailable (EVA, comms blackout, illness). To prevent single points of failure, each critical controller declares a primary owner and a backup; the system should hand over control seamlessly and log the event.

For asset $a$, denote the primary controller $o(a)$ and backup $b(a)$. Let availability random variables $A_{o},A_{b}\in\{0,1\}$ indicate whether the controller is online. The control function $c(a)$ is
\[
c(a)=
\begin{cases}
o(a), & A_{o}=1,\\
b(a), & A_{o}=0\ \wedge\ A_{b}=1,\\
\varnothing, & A_{o}=0\ \wedge\ A_{b}=0,
\end{cases}
\]
with a \emph{role-switch} event when $c(a)=b(a)$. The runtime samples availability from an outage rate $p\in[0,1]$ and triggers handoff; switches and residual failures are counted for evaluation and AMPI. Ablation results of switching on or off dynamic redundancy is reported in Table \ref{tab:redundancy}.

Assuming independence,
\[
\Pr(\text{serviceable }a)=1-\Pr(A_{o}=0, A_{b}=0)=1-p_o\,p_b,\qquad
\mathbb{E}[\text{switches}]=\sum_{a\in\mathcal{A}} \Pr(A_{o}=0, A_{b}=1),
\]
and for homogeneous $p_o=p_b=p$ across $|\mathcal{A}|$ assets, $\mathbb{E}[\text{switches}]=|\mathcal{A}|\,p(1-p)$. These quantities appear in the metrics as \textit{RoleSw} and \textit{Failures}.

% Figure idea: a small state diagram for one asset showing Owner-online, Switch-to-Backup, and Failure states with probabilities p and (1-p); overlay a bar showing RoleSw counts across assets.
% Table idea: per-asset Owner/Backup and observed (switches, failures) over a run; caption "Observed reliability under outage rate p".

\paragraph{Functional distributed leadership.}
A single permanent leader simplifies escalation but may bottleneck specialization; conversely, allowing leadership to follow function (ops vs.\ emergency vs.\ science) can reduce decision latency and improve task fit.

Define a leader selection map $L:\{\textsc{DailyOps},\textsc{Emergency},\textsc{Science}\}\times\{\texttt{single},\texttt{functional}\}\to \mathcal{V}$:
\[
L(s,m)=
\begin{cases}
\mathrm{CMD}, & m=\texttt{single},\\
\mathrm{OPS}, & m=\texttt{functional}\ \wedge\ s=\textsc{DailyOps},\\
\mathrm{CMD}, & m=\texttt{functional}\ \wedge\ s=\textsc{Emergency},\\
\mathrm{GEO}\ \text{or}\ \mathrm{BIO}, & m=\texttt{functional}\ \wedge\ s=\textsc{Science},\\
\mathrm{CMD}, & \text{otherwise},
\end{cases}
\]
with deterministic tie-breaking via the roster. The selected leader seeds the agenda/plan, anchors directives, and resolves undecided consensus. The mode (\texttt{single} vs.\ \texttt{functional}) and scenario are runtime parameters; the choice is logged and reported alongside routing policy to support A/B studies.

Let $D(L)$ denote the expected communication diameter from leader $L$ to the set of executors for a scenario; a functional leader that sits closer in the hierarchy to the working set can lower $D(L)$, reducing expected messages/time and improving AMPI components for time and messages. This is validated by comparing the \texttt{single} vs.\ \texttt{functional} modes under identical seeds and rosters.

The communications policy switch and whitelist, hub-forwarding, translator/dialect normalizer, and cross-layer accounting are enforced by the router and message loop; the leader selector applies the $L(s,m)$ mapping; the asset controller performs owner/backup handoff and logs \emph{RoleSw}; and the metrics module exposes $\rho_{\text{cross}}$, \emph{Failures}, \emph{Msgs}, \emph{Time}, which feed AMPI via monotone squashing $x\mapsto x/(x+K)$ per component. These are surfaced in per-run summaries and can be exported to CSV for statistical aggregation.

\subsection{Innovative Interaction Strategies for Multi-Agent Collaboration}
\label{subsec:interaction}

Beyond structural routing, Agent Mars equips agents with interaction strategies that explicitly model memory, consensus, and communication heterogeneity. These are motivated by real mission needs: situational awareness over time, agreement under conflict, and translation across disciplinary jargon. Each mechanism is implemented in the runtime loop and exposed to experiments for A/B evaluation.

\paragraph{Scenario-aware memory.}

\begin{figure}[t]
\centering
    \includegraphics[width=\linewidth]{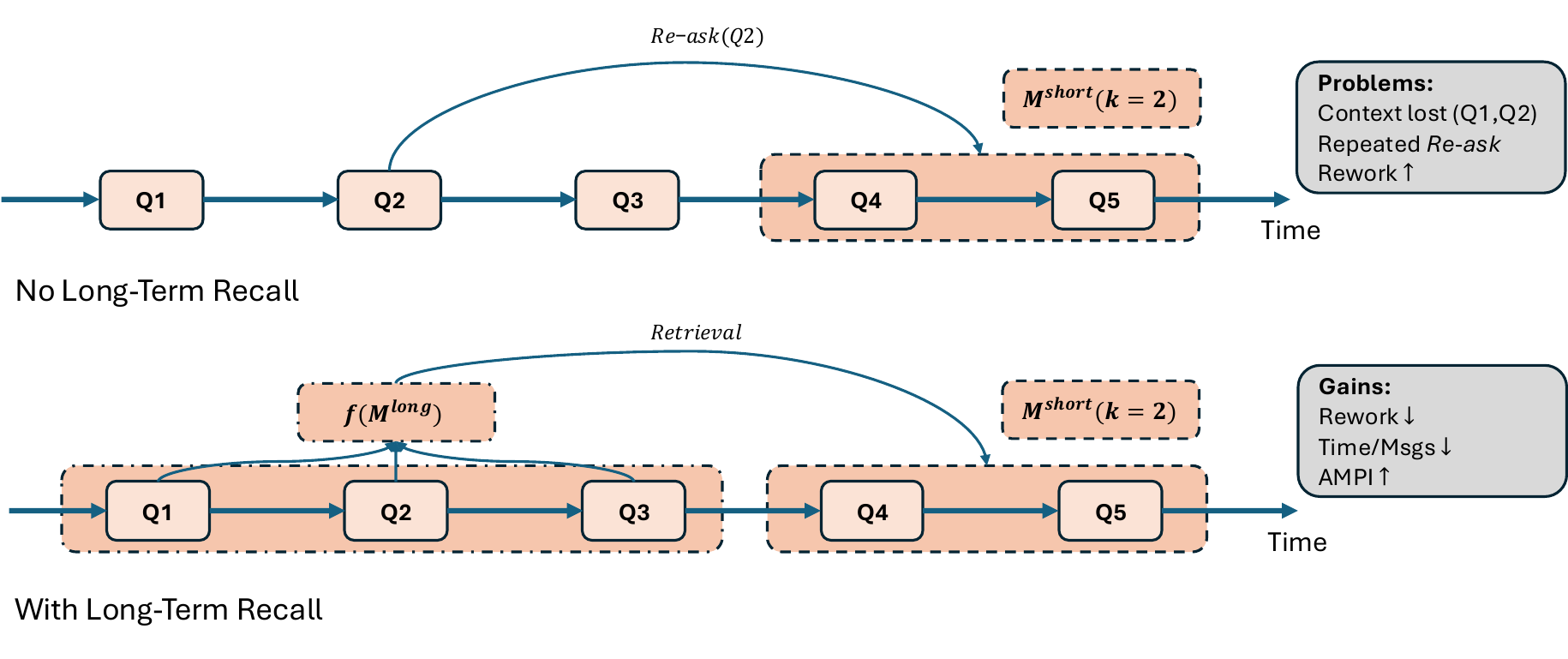}
\caption{Scenario-aware memory timeline. Top: short-term window only ($M^{\text{short}}$ with horizon $k$) forgets early context (\(Q1\)–\(Q2\)), so later queries re‑ask earlier items (e.g., \(Q2 \rightarrow Q5\)). Bottom: adding distilled long‑term recall \(f(M^{\text{long}})\) (and optional shared memory) enables state carry‑over, reducing re‑asks and rework; see Table~\ref{tab:memory}.}
\label{fig:memory-timeline}
\end{figure}

LLM agents without memory treat every utterance in isolation, which undermines continuity. Real missions, however, require both short-term recall (current task state) and long-term recall (role experience, prior incidents).  
We model each agent $i$ with two memory buffers: a short-term window $M^{\mathrm{short}}_i$ that stores the last $k$ turns, and a long-term store $M^{\mathrm{long}}_i$ of distilled past events. Shared-memory modes let selected agents access a common pool $M^{\mathrm{shared}}$, while isolated modes enforce privacy.

Formally, let $q_t$ be the query at time $t$. The retrieval context is
\[
C_i(t) = \mathrm{concat}\big( M^{\mathrm{short}}_i(t),\; f(M^{\mathrm{long}}_i),\; M^{\mathrm{shared}}(t)\mathbf{1}_{\text{shared}}\big),
\]
where $f(\cdot)$ is a summarization operator and $\mathbf{1}_{\text{shared}}$ toggles shared mode. Ablations compare off/basic/shared settings (Table~\ref{tab:memory}).

\paragraph{Consensus formation.}

Task conflicts (e.g., EVA scheduling under resource contention) require agents to negotiate rather than unilaterally execute. Inspired by classical distributed consensus, we implement a lightweight propose–vote cycle: agents broadcast proposals $p_i$, then vote until quorum is reached or rounds expire.

Let $P=\{p_1,\dots,p_n\}$ be proposals and $V(p_i)$ the votes received. Consensus is declared if
\[
\frac{|V(p_i)|}{n} \;\geq\; \theta,
\]
where $\theta\in(0,1]$ is the quorum threshold. Runtime parameters control the number of rounds $R$ and $\theta$. Default is $R=2$, $\theta=0.6$ (Table~\ref{tab:factors}).  

\noindent\textbf{Consensus diagnostics.}
Let $r=1,\dots,R$ denote debate rounds. At round $r$, define the vote share for proposal $p\in P$ as
\[
s_r(p)=\frac{|V_r(p)|}{n},
\]
where $V_r(p)$ is the set of votes received by $p$ at round $r$.

\emph{Time-to-consensus} is the first round that reaches quorum:
\[
r^\star=\min\Big\{r:\max_{p\in P}s_r(p)\ge \theta\Big\},
\]
and we set $r^\star=R{+}1$ if no proposal reaches quorum.

\emph{Vote divergence} is quantified by (i) normalized vote entropy
\[
D_r=-\frac{1}{\log|P|}\sum_{p\in P}s_r(p)\log\!\big(s_r(p)+\varepsilon\big),
\qquad \varepsilon=10^{-12},
\]
and (ii) the top-1 margin
\[
\Delta_r=s_r(p_{(1)})-s_r(p_{(2)}),
\]
where $p_{(1)}$ and $p_{(2)}$ are the most- and second-most-voted proposals at round $r$.

\emph{Downstream task quality} is scored against the scenario deliverables listed in Table~\ref{tab:scenario-prompts}. Let a scenario specify $J$ deliverables; we define a checklist score
\[
Q=\frac{1}{J}\sum_{j=1}^{J}\mathbf{1}\{\text{deliverable}_j\ \text{is satisfied}\}-\lambda\,N_{\mathrm{viol}},
\]
where $N_{\mathrm{viol}}$ counts explicit constraint violations (e.g., safety interlocks, redlines) recorded in the run log, and $\lambda\ge 0$ controls penalty strength (default $\lambda=0$).

\paragraph{Heterogeneous technical-language protocols.}

Different specialist groups (e.g., geology, medicine, power systems) use distinct technical dialects. Direct exchanges across groups risk misinterpretation. To capture this, we assign each group a domain-specific protocol template and interpose a translator agent.  

Formally, let $\mathcal{L}_g$ denote the lexicon/protocol for group $g$. A message $m\in\mathcal{L}_g$ destined for group $g'$ is rewritten as
\[
m' = \mathrm{Translate}(m;\; \mathcal{L}_g \to \mathcal{L}_{g'}),
\]
with the translator agent mediating. This incurs additional message overhead but improves comprehension.  
Ablations toggle protocols=off vs.\ hetero (Table~\ref{tab:protocols}), measuring whether translation overhead is offset by reduced failure/miscommunication.

\begin{figure}[t]
\centering
\includegraphics[width=\linewidth]{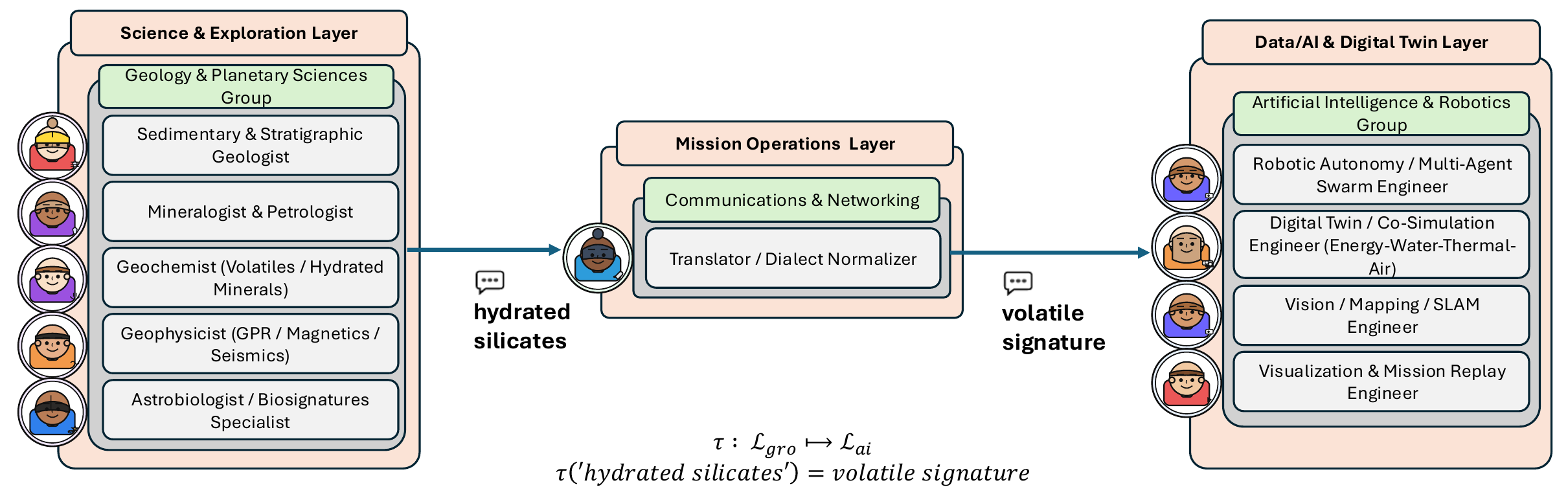}
\caption{Heterogeneous protocols with a translator agent. GEO issues a domain term
\texttt{hydrated silicates}; the Translator maps it via
$\tau:\mathcal{L}_{\text{geo}}\!\to\!\mathcal{L}_{\text{ai}}$
into AI’s control vocabulary \texttt{volatile signature}, recording an audit trail.}
\label{fig:translator-cartoon}
\end{figure}

\begin{figure}[htbp]
    \centering
    \includegraphics[width=\linewidth]{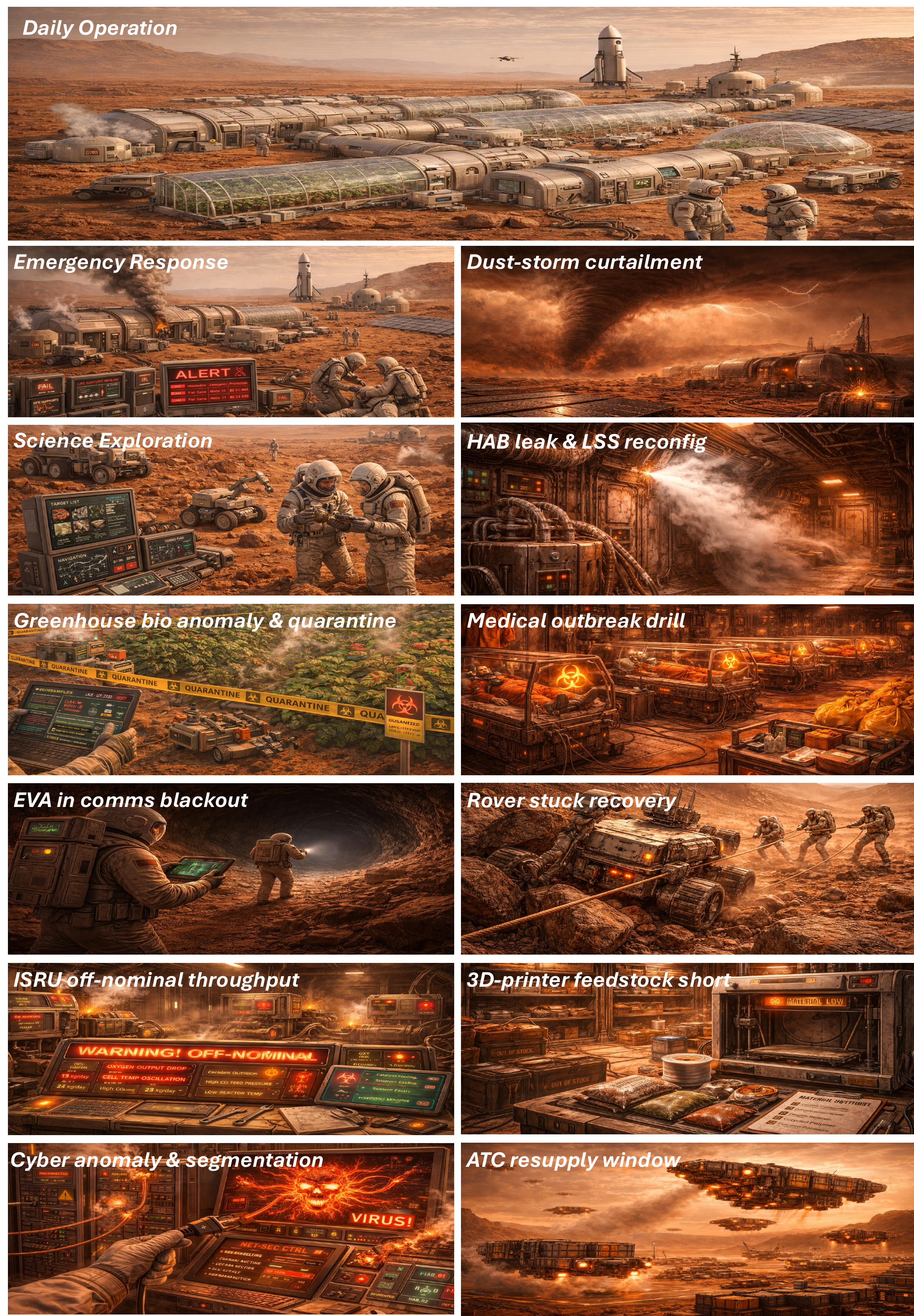}
    \caption{Conceptual illustrations of the 13 scenarios in the Mars Benchmark Task Suite (see Table~\ref{tab:scenario-prompts}).}
    \label{fig:scenarios}
\end{figure}

Together, these strategies allow Agent Mars to study not only structural routing but also the micro-dynamics of how agents remember, agree, and understand each other under mission stress.

\subsection{Mars Benchmark Task Suite and Agent Mars Performance Index (AMPI)}
\label{subsec:bench-ampi}

\paragraph{Benchmark Task Suite}

\paragraph{Seed prompts.}
For each scenario we fix a single seed prompt that is held constant across all routing/leadership/memory/consensus/protocol settings; only the runtime toggles change. The exact seeds are listed in Table~\ref{tab:scenario-prompts}.

\begin{longtable}{@{}p{0.25\linewidth}p{0.73\linewidth}@{}}
\caption{Scenario prompts for the Mars Benchmark Task Suite (seed text given to the runner).}
\label{tab:scenario-prompts}\\
\toprule
\textbf{Short title} & \textbf{Scenario prompt (seed to the runner)}\\
\midrule
\endfirsthead
\toprule
\textbf{Short title} & \textbf{Scenario prompt (seed to the runner)}\\
\midrule
\endhead
\bottomrule
\endfoot

Daily operations &
You are running the \textsc{Daily Operations} script for the next 8\,h on Mars. \textbf{Trigger:} start‑of‑sol stand‑up. \textbf{Tasks:} roll up overnight telemetry, set priorities, allocate power/water/crew time, schedule EVA/ROV/ISRU jobs, confirm asset heartbeats (\texttt{HAB}, \texttt{SOL\_CTRL}, \texttt{NUKE\_CTRL}, \texttt{ISRU\_PLANT}, \texttt{GH\_CTRL}, \texttt{ROV\_SCI/INSP/CARGO}, \texttt{UAV\_MAP/UAV\_COM}, \texttt{PRT\_CTRL}, \texttt{ARM\_CTRL}). \textbf{Constraints:} intermittent comms to Earth; limited battery reserve; obey safety interlocks. \textbf{Deliverables:} (1) 8‑h agenda with owners and time tags; (2) resource allocation summary; (3) risks \& abort thresholds; (4) end‑of‑run status note. \\

Emergency response &
You are running the \textsc{Emergency Response} script after a compound alert. \textbf{Trigger:} microgrid voltage sag, elevated \texttt{LSS} CO$_2$ scrubber load, and a transient suit sensor glitch reported by \texttt{SUIT\_CTRL}. \textbf{Tasks:} stabilize power, verify life support margins, rule out suit hardware fault, replan near‑term ops. \textbf{Constraints:} treat any redline as authority‑to‑stop; Earth unavailable. \textbf{Deliverables:} (1) 30‑min stabilization plan; (2) incident command structure (who leads what); (3) prioritized task list with timers; (4) return‑to‑nominal criteria \& handoff. \\

Science exploration &
You are running the \textsc{Science Exploration} script for a 3‑hour traverse. \textbf{Trigger:} GEO requests sampling along the Aeolis rim. \textbf{Tasks:} propose traverse, sample targets and chain‑of‑custody, COM link windows, AI/SLAM checks; prep \texttt{ROV\_SCI}, \texttt{UAV\_MAP}. \textbf{Constraints:} sunlight/thermal limits; sparse relay from \texttt{COMSAT\_CTRL}. \textbf{Deliverables:} (1) route with waypoints \& hazards; (2) sample list with rationale \& custody steps; (3) comms plan; (4) go/no‑go with aborts and a contingency route. \\

Greenhouse bio anomaly \& quarantine &
You are running the \textsc{Greenhouse Bio Anomaly} script. \textbf{Trigger:} \texttt{GH\_CTRL} flags leaf spotting and elevated spore counts in Zone~2. \textbf{Tasks:} BIO diagnosis, AGRI ops changes, LSS circulation implications, quarantine intensity debate if needed. \textbf{Constraints:} protect crew \& food supply; avoid cross‑contamination. \textbf{Deliverables:} (1) quarantine plan (zones, PPE, duration); (2) sampling \& lab tests using \texttt{LAB\_INSTR}; (3) greenhouse setpoint changes; (4) success/clear criteria and monitoring cadence. \\

EVA in comms blackout &
You are running the \textsc{EVA in Comms Blackout} script. \textbf{Trigger:} planned EVA route crosses a 20‑minute RF shadow. \textbf{Tasks:} COM link‑budget and relay plan (\texttt{UAV\_COM} loiter), store‑and‑forward procedures, timed callouts, hazard review with GEO. \textbf{Constraints:} hard comms blackout segment; suit consumables. \textbf{Deliverables:} (1) EVA timeline; (2) comms plan (waypoints, altitudes, timing); (3) emergency actions for loss of relay; (4) go/no‑go with abort thresholds. \\

ISRU off‑nominal throughput &
You are running the \textsc{ISRU Off‑Nominal Throughput} script. \textbf{Trigger:} oxygen output fell from 24\,kg/day to 16\,kg/day; cell temperature oscillation observed at \texttt{ISRU\_PLANT}. \textbf{Tasks:} triage (sensor vs.\ process), hypothesize catalyst fouling or feedstock moisture, coordinate PWR peaks, plan derate. \textbf{Constraints:} do not exceed thermal redlines; maintain base O$_2$ buffer. \textbf{Deliverables:} (1) root‑cause hypotheses and quick tests; (2) safe derate schedule; (3) spares/assay requests to \texttt{LAB\_INSTR}; (4) return‑to‑service checklist. \\

Cyber anomaly \& segmentation &
You are running the \textsc{Cyber Anomaly \& Segmentation} script. \textbf{Trigger:} \texttt{NET\_SEC\_CTRL} reports unusual outbound traffic from \texttt{HAB\_02} at 02{:}17. \textbf{Tasks:} triage (malware vs.\ misconfig), apply least‑privilege segmentation, update allowlist, preserve logs for forensics. \textbf{Constraints:} keep life‑critical systems online; audit every action. \textbf{Deliverables:} (1) incident timeline; (2) containment/segmentation actions; (3) impact assessment; (4) recovery plan \& monitoring. \\

Dust‑storm curtailment &
You are running the \textsc{Dust‑Storm Curtailment} script. \textbf{Trigger:} \texttt{WX\_STATION} forecasts storm in 18\,h; optical depth rising 1.2$\rightarrow$2.8. \textbf{Tasks:} curtail non‑critical loads, set cleaning \& stow policies for \texttt{SOL\_CTRL}/\texttt{UAV\_MAP}, reserve battery for \texttt{LSS} and comms, staffing plan. \textbf{Constraints:} limited storage; reduced solar input. \textbf{Deliverables:} (1) curtailment schedule \& priorities; (2) energy budget; (3) crew plan; (4) restart criteria post‑storm. \\

HAB leak \& LSS reconfig &
You are running the \textsc{HAB Leak \& LSS Reconfiguration} script. \textbf{Trigger:} pressure drop 0.2\,kPa/min in \texttt{HAB\_01} Ring~B. \textbf{Tasks:} localize leak (with \texttt{ROV\_INSP} or crew), isolate segment, reconfigure LSS loops, prep patch. \textbf{Constraints:} preserve habitable zones; \texttt{AIRLOCK\_CTRL} interlocks. \textbf{Deliverables:} (1) isolation plan and commands; (2) leak‑locate procedure; (3) patch \& verification steps; (4) dwell time limits and all‑clear criteria. \\

Medical outbreak drill &
You are running the \textsc{Medical Outbreak Drill} (infectious disease). \textbf{Trigger:} two crew with fever/cough and one GI case after shared meal. \textbf{Tasks:} triage, isolation, contact tracing, ward workflow (MED/NUR), PSY crew support, logistics for PPE and meds. \textbf{Constraints:} avoid clinic overload; maintain critical ops. \textbf{Deliverables:} (1) case definition and cohorting; (2) ward/PPE plan; (3) treatment \& testing algorithm; (4) clearance \& return‑to‑work rules; (5) comms to base. \\

Rover stuck recovery &
You are running the \textsc{Rover Stuck Recovery} script. \textbf{Trigger:} \texttt{ROV\_SCI} reports slip ratio 0.8 on fine regolith, slope 8°, battery 44\%. \textbf{Tasks:} assess terrain with GEO/AI, egress options (backdrive, anchor, traction aids), \texttt{UAV\_MAP} overflight if useful, protect battery SOC. \textbf{Constraints:} avoid wheel trenching; time‑boxed attempts. \textbf{Deliverables:} (1) recovery plan with stepwise commands; (2) abort/assist criteria; (3) comms windows; (4) post‑recovery inspection checklist. \\

3D‑printer feedstock short &
You are running the \textsc{3D‑Printer Feedstock Short} script. \textbf{Trigger:} \texttt{PRT\_CTRL} inventory shows 0.6\,kg engineering filament remaining vs.\ 1.8\,kg needed within 48\,h for valve clamps and a manifold ring. \textbf{Tasks:} evaluate substitutes (recycled HDPE, basalt‑fiber regolith composite, sintered regolith) with LOGT/ISRU/LAB; define print/process settings and qualification tests. \textbf{Constraints:} mechanical/thermal spec, outgassing limits, contamination control. \textbf{Deliverables:} (1) candidate ranking with pros/cons; (2) BOM \& print parameters; (3) test protocol (strength, thermal cycle, off‑gassing); (4) production schedule with risk/mitigations. \\

ATC resupply window &
You are running the \textsc{ATC Resupply Window} script. \textbf{Trigger:} two vehicles request overlapping landing slots; EVA drone ops nearby. \textbf{Tasks:} allocate \texttt{ATC\_LZ} windows, COM air‑ground timeslots, ground handling flow with LOGT, no‑fly zones for \texttt{UAV\_COM/UAV\_MAP}. \textbf{Constraints:} safety first; fixed approach corridor; limited ground crew. \textbf{Deliverables:} (1) slot schedule \& right‑of‑way; (2) comms plan; (3) ground handling timeline; (4) reserves/abort windows and notification script. \\
\end{longtable}

% =======================
% Failure definition block
% Place this right after Table~\ref{tab:scenario-prompts} and BEFORE
% \paragraph{Agent Mars Performance Index (AMPI).}
% =======================

\paragraph{Failure events and accounting.}\label{par:failures}
To remove ambiguity, we decompose \textit{Failures} into three measurable types and
report their (optionally aggregated) counts from the run log.

\begin{center}
\fbox{\begin{minipage}{0.97\linewidth}
\small
\textbf{Failure types (one run).}
Let $\mathcal{A}$ be the set of asset controllers, and let $\tau$ denote the full
run transcript/log (messages + system events) for scenario $s$.
Each scenario specifies a deliverable set $\mathcal{D}_s=\{d_j\}_{j=1}^{J}$
(Table~\ref{tab:scenario-prompts}).

\begin{itemize}
\item \textbf{Type A: asset unserviceable ($N_{\mathrm{asset}}$).}
For each asset $a\in\mathcal{A}$, the runtime selects a controller
(primary if online; else backup if online; else none).
We count an asset as \emph{unserviceable} if it becomes uncontrollable at any time:
\[
N_{\mathrm{asset}} \;=\; \sum_{a\in\mathcal{A}}
\mathbf{1}\Big\{\exists t:\; c_t(a)=\varnothing \Big\},
\]
where $c_t(a)\in\mathcal{V}\cup\{\varnothing\}$ is the selected controller at time $t$.
(Thus, repeated outages of the \emph{same} asset within a run are counted once.)

\item \textbf{Type B: constraint violation ($N_{\mathrm{viol}}$).}
We log a violation event whenever an action (or proposed plan) breaches a
scenario constraint or a safety rule, including (non-exhaustive):
(i) safety \emph{redlines} (authority-to-stop conditions),
(ii) interlock bypass/override attempts (e.g., \texttt{AIRLOCK\_CTRL} or habitat pressure locks),
(iii) execution of a step explicitly rejected by the Safety \& Ethics Officer.
Let $\mathcal{E}_{\mathrm{viol}}(\tau)$ be the set of such logged events; then
\[
N_{\mathrm{viol}} \;=\; \big|\mathcal{E}_{\mathrm{viol}}(\tau)\big|.
\]
(Repeated violations are counted separately if they correspond to distinct logged events.)

\item \textbf{Type C: deliverable missing ($N_{\mathrm{miss}}$).}
Each scenario has $J$ required deliverables $\{d_j\}_{j=1}^{J}$.
A deterministic checker $\phi_j(\tau)\in\{0,1\}$ marks whether deliverable $d_j$ is satisfied
from the final run output and log (e.g., presence of required sections/fields).
We count missing deliverables as
\[
N_{\mathrm{miss}} \;=\; \sum_{j=1}^{J} \Big(1-\phi_j(\tau)\Big).
\]
\end{itemize}

\textbf{What we report in tables.}
Unless stated otherwise, the \textit{Failures} column reports the aggregated failure count
\[
F \;=\; N_{\mathrm{asset}} \;+\; N_{\mathrm{viol}} \;+\; N_{\mathrm{miss}}.
\]
\end{minipage}}
\end{center}

\paragraph{Agent Mars Performance Index (AMPI).}
To support single‐number comparisons while preserving interpretability, we define
\begin{equation}
\mathrm{AMPI}
= w_1\,(1-\tilde{T}) + w_2\,(1-\tilde{M}) + w_3\,(1-\tilde{C}) + w_4\,(1-\tilde{F}) + w_5\,(1-\tilde{S}),
\label{eq:ampi}
\end{equation}
where $\tilde{T},\tilde{M},\tilde{C},\tilde{F},\tilde{S}\in[0,1]$ are normalized time, messages, cross‐layer ratio, failures, and role switches, respectively. For non‐ratio terms we apply a monotone squashing
\begin{equation}
\tilde{X}=\frac{X}{X+K_X},\qquad X\in\{T,M,F,S\},
\label{eq:squash}
\end{equation}
with positive constants $K_T,K_M,K_F,K_S$.

\paragraph{Time measurement.}
In our implementation, $T$ denotes the end-to-end wall-clock runtime of a scenario run, capturing both dialog length and external API/network latency. To support latency-robust comparisons, we additionally report surrogate effort measures that can replace $T$ in Eq.~\eqref{eq:squash}, such as the number of LLM calls $N_{\mathrm{call}}$, the number of agent decision steps $N_{\mathrm{step}}$, or total token usage $N_{\mathrm{tok}}$. When using a surrogate, we set $T\leftarrow N_{\mathrm{call}}$ (or $N_{\mathrm{step}},N_{\mathrm{tok}}$) and choose the corresponding $K_T$.

By default we use $\mathbf{w}=[0.4,\,0.2,\,0.0,\,0.25,\,0.15]$, and exclude the cross‐layer penalty unless \texttt{--ampi\_include\_crosslayer true} is set (in which case $w_3$ is honored).

% ------------------------------------------------------------------
% Mars Benchmark Task Suite (13 scenarios) -- no ID column
% ------------------------------------------------------------------
\begin{longtable}{@{}p{0.25\linewidth}p{0.20\linewidth}p{0.55\linewidth}@{}}
\caption{Mars Benchmark Task Suite: scenarios at a glance.}
\label{tab:scenarios}\\
\toprule
\textbf{Short title} & \textbf{Leader (order)} & \textbf{Focus / main participants (\& assets)}\\
\midrule
\endfirsthead
\toprule
\textbf{Short title} & \textbf{Leader (order)} & \textbf{Focus / main participants (\& assets)}\\
\midrule
\endhead
\bottomrule
\endfoot

Daily operations
& OPS $\succ$ CMD
& Morning agenda; cross-team check-ins across MED, NUR, LSS, PWR, ISRU, AGRI, GEO, EVA, COM, LOGT, MNT, DKM, PSY; energy allocation; optional GEO$\rightarrow$AI planning; asset heartbeats (\textit{HAB, SOL\_CTRL, NUKE\_CTRL, ISRU\_PLANT, GH\_CTRL, ROV\_SCI/INSP/CARGO, UAV\_MAP/UAV\_COM, PRT\_CTRL, ARM\_CTRL}).\\

Emergency response
& CMD $\succ$ OPS
& Engineering diagnostics (LSS, PWR, ISRU, MNT); medical triage (MED, NUR); logistics/ops replanning (LOGT, OPS); science/environment risk (GEO, DKM); asset auto-protect (\textit{HAB, SOL\_CTRL, NUKE\_CTRL, ISRU\_PLANT, GH\_CTRL}).\\

Science exploration
& GEO $\succ$ BIO $\succ$ OPS
& Science plan; GEO$\rightarrow$AI navigation/SLAM checks; GEO$\rightarrow$COM link-budget assessment; rover pre-check (\textit{ROV\_SCI}).\\

Greenhouse bio anomaly \& quarantine
& AGRI $\succ$ BIO $\succ$ OPS
& BIO diagnosis; AGRI ops changes; LSS impact on circulation/filters; asset: \textit{GH\_CTRL}. Consensus on quarantine intensity (AGRI, BIO, OPS, PSY).\\

EVA in comms blackout
& OPS $\succ$ COM $\succ$ GEO
& GEO EVA route \& risks; COM link budget \& blind-spot mitigation; AI SLAM keyframe checks; asset: \textit{UAV\_COM} as relay. Consensus on EVA bandwidth split (COM, OPS, AI).\\

ISRU off-nominal throughput
& ISRU $\succ$ PWR $\succ$ OPS
& ISRU root-cause hypotheses; PWR supply/peaks check; LAB catalyst/chemistry assays; asset: \textit{ISRU\_PLANT} safe derate.\\

Cyber anomaly \& segmentation
& COM $\succ$ OPS $\succ$ DKM
& COM detection/whitelist updates; DKM threat hypotheses; AI node integrity checks; (optional) consensus on network segmentation levels (COM, OPS, AI, DKM).\\

Dust-storm curtailment
& PWR $\succ$ OPS
& DKM storm forecast; PWR curtailment plan (shed priorities); OPS staffing/guard strategy; assets: \textit{SOL\_CTRL}, \textit{NUKE\_CTRL}.\\

HAB leak \& LSS reconfig
& LSS $\succ$ OPS $\succ$ MNT
& MNT leak localization; LSS loop/pressure/filter reconfiguration; asset: \textit{HAB} segment isolation and $\Delta P$ control.\\

Medical outbreak drill
& MED $\succ$ OPS $\succ$ NUR
& MED triage/isolation; NUR PPE/ward workflow; PSY crew support; (no assets).\\

Rover stuck recovery
& OPS $\succ$ GEO $\succ$ MNT
& GEO terrain risk/egress route; AI outcome simulations; MNT traction/anchor/recovery plan; assets: \textit{ROV\_SCI}, \textit{ROV\_INSP}.\\

3D-printer feedstock short
& LOGT $\succ$ ISRU $\succ$ OPS
& LOGT stocktake \& rationing; ISRU in-situ alternative feed prep; LAB tests (strength/volatiles/contamination); assets: \textit{PRT\_CTRL}, \textit{ARM\_CTRL}.\\

ATC resupply window
& ATC $\succ$ OPS $\succ$ COM
& ATC landing slots \& conflict resolution; COM air-ground link timeslots; LOGT ground handling/flows; asset: \textit{UAV\_COM}. Consensus on slot/bandwidth \& ground right-of-way (ATC, OPS, COM, LOGT).\\

\end{longtable}

\section{Results}
\label{sec:results}

\subsection{Experimental setup and implementation details}
\label{subsec:setup}

All experiments are executed with the public runner (\emph{main.py}) using ChatGPT as the LLM backend. We use deterministic decoding (temperature $=0$) to reduce stochasticity in dialog trajectories, so differences across configurations primarily reflect routing/leadership/memory/consensus/protocol design choices. Each configuration is repeated for $N=20$ runs and we report mean values in the tables. The wall-clock time metric includes external API/network latency; therefore, we also report message counts and failure-related counters to provide latency-robust evidence of coordination effort. Failures are computed as the aggregated failure count $F$ defined in \S\ref{par:failures},
with a recorded breakdown $(N_{\mathrm{asset}},N_{\mathrm{viol}},N_{\mathrm{miss}})$ for diagnostics.

Unless a factor is being ablated, the default setting is: routing = STRICT, leadership = functional, role switching = on, memory = shared, consensus = off, and heterogeneous protocols = off. AMPI follows Eq.~\eqref{eq:ampi} with the default constants and weights in \S\ref{subsec:bench-ampi}. The cross-layer term is disabled by default ($w_3=0$), but the cross-layer utilization ratio is still reported for transparency.

\subsection{Factors and levels}
\label{subsec:results-factors}

Table~\ref{tab:factors} summarizes the experimental factors and levels. Each sweep varies one factor while holding the others at their defaults, enabling attribution of changes to a single design decision.

\subsection{Core comparison: STRICT vs.\ CROSSLAYER routing; single vs.\ functional leadership}
\label{subsec:results-core}

Table~\ref{tab:core-ab} reports the main end-to-end comparison across routing (STRICT, CROSSLAYER) and leadership (single, functional), with other modules fixed to their defaults.

Across planning- and communication-heavy scenarios, curated cross-layer routing consistently reduces end-to-end time and often reduces message count. For example, in DailyOperations, moving from STRICT/single to CROSSLAYER/functional reduces time from 232.4\,s to 191.9\,s (17.4\% reduction) with a small reduction in messages (43 to 42) and a higher measured cross-layer utilization (0.00 to 0.10). ScienceExploration exhibits the largest benefit from cross-layer links: STRICT/single takes 61.2\,s versus 29.8\,s under CROSSLAYER/functional (51.3\% reduction), with fewer messages (7 to 5) and substantially higher cross-layer utilization (0.00 to 0.45). CommsBlackoutEVA similarly improves from 58.3\,s (STRICT/single) to 41.8\,s (CROSSLAYER/functional), a 28.3\% reduction, with messages decreasing from 12 to 11.

Functional leadership generally improves over single leadership when the scenario’s work is concentrated in a specific functional layer, consistent with reduced escalation overhead. This is visible in multiple STRICT rows, such as ScienceExploration (61.2\,s to 56.0\,s), CommsBlackoutEVA (58.3\,s to 52.9\,s), and CyberAnomaly (37.9\,s to 34.1\,s). EmergencyResponse shows minimal sensitivity to both routing and leadership (296.2\,s to 292.9\,s), consistent with hub-centric flows that already route through command/operations.

Not all scenarios benefit equally from cross-layer exchanges. DustStormCurtail shows a small slowdown under CROSSLAYER/functional relative to STRICT/functional (45.9\,s to 50.8\,s), suggesting that additional cross-layer coordination can be overhead-dominant when the task is already well-structured. GH\_BioOutbreak and several shorter scenarios show smaller differences, indicating that gains depend on whether the scenario induces repeated cross-team dependency resolution.

Overall, AMPI changes track time and failure differences under the default weighting, while the reported cross-layer ratio clarifies when speed improvements are achieved via heavier cross-layer use.

\subsection{Reliability ablation: dynamic role switching}
\label{subsec:results-switch}

Table~\ref{tab:redundancy} evaluates redundancy by toggling role switching on versus off under STRICT routing. Enabling role switching reduces both time and failures across most scenarios, with the largest improvements in longer runs where asset availability is a frequent bottleneck. DailyOperations improves from 224.8\,s (off) to 188.3\,s (on), and EmergencyResponse improves from 338.1\,s to 298.6\,s. Failure counts consistently decrease when role switching is enabled (e.g., DailyOperations: 0.50 to 0.12; EmergencyResponse: 0.70 to 0.25), indicating that automated handover prevents stalled control paths. As expected, RoleSw increases in the on condition, reflecting active failover events. AMPI improves modestly in most cases because the reduction in time and failures outweighs the penalty associated with additional role switches.

\subsection{Memory sweep: off vs.\ basic vs.\ shared}
\label{subsec:results-memory}

Table~\ref{tab:memory} compares memory modes under \texttt{CROSSLAYER} routing and functional
leadership. Overall, memory affects both efficiency and robustness, and the effect is not monotonic:
\texttt{basic} memory often yields the lowest runtime, whereas \texttt{shared} memory can reduce
failures in longer, multi-step scenarios but may introduce retrieval noise in compact workflows.

In \textsc{DailyOperations}, \texttt{basic} memory reduces time from 198.0\,s to 165.0\,s, while
\texttt{shared} memory yields 191.9\,s but substantially reduces failures (0.20 to 0.05) and increases
AMPI (0.51 to 0.54). A similar pattern is observed in \textsc{EmergencyResponse} (311.0\,s to
281.0\,s for \texttt{basic}; 292.9\,s for \texttt{shared}, with failures 0.60 to 0.10).
\texttt{basic} memory also provides the largest speedups in scenarios such as \textsc{GH\_BioOutbreak}
(51.3\,s to 44.0\,s) and \textsc{DustStormCurtail} (50.8\,s to 46.3\,s), whereas \texttt{shared} returns
closer to the \texttt{off} condition in these cases (e.g., 50.4\,s and 50.8\,s, respectively).

Notably, \texttt{shared} memory yields clear efficiency gains in coordination-heavy scripts such as
\textsc{CyberAnomaly} (40.4\,s to 31.7\,s), \textsc{CommsBlackoutEVA} (45.6\,s to 41.8\,s), and
\textsc{ATC\_ResupplyWindow} (42.4\,s to 32.9\,s). In contrast, \texttt{shared} memory can be
overhead-dominant in tightly scoped workflows, e.g., \textsc{ISRU\_OffNominal} (21.7\,s to 27.1\,s)
and \textsc{RoverStuckRecovery} (41.8\,s to 44.0\,s). Short, low-dependency scenarios show smaller
sensitivity to memory mode (e.g., \textsc{ScienceExploration}: 29.6\,s, 29.1\,s, and 29.8\,s).

\subsection{Consensus formation: propose--vote}
\label{subsec:results-consensus}

Table~\ref{tab:consensus2} evaluates the propose--vote consensus mechanism under
\texttt{CROSSLAYER} routing with shared memory and functional leadership.
As designed, enabling consensus increases coordination traffic across all tested scenarios
(e.g., \textsc{DailyOperations}: 42$\rightarrow$48 messages; \textsc{ScienceExploration}: 5$\rightarrow$9).

The effect on runtime is mixed and scenario-dependent. Consensus yields a modest speedup in
\textsc{DailyOperations} (191.9\,s $\rightarrow$ 184.0\,s) and a small improvement in
\textsc{ATC\_ResupplyWindow} (32.9\,s $\rightarrow$ 32.1\,s), but introduces slowdowns in
\textsc{EmergencyResponse} (292.9\,s $\rightarrow$ 307.0\,s),
\textsc{ScienceExploration} (29.8\,s $\rightarrow$ 35.1\,s),
\textsc{GH\_BioOutbreak} (50.4\,s $\rightarrow$ 58.8\,s), and
\textsc{CommsBlackoutEVA} (41.8\,s $\rightarrow$ 46.4\,s).

Reliability effects are similarly heterogeneous. Consensus substantially reduces the aggregated
failure count in \textsc{EmergencyResponse} (0.10 $\rightarrow$ 0.01) and slightly in
\textsc{GH\_BioOutbreak} (0.02 $\rightarrow$ 0.01), while increasing failures in
\textsc{DailyOperations} (0.05 $\rightarrow$ 0.30) and \textsc{ScienceExploration}
(0.01 $\rightarrow$ 0.20); \textsc{CommsBlackoutEVA} and \textsc{ATC\_ResupplyWindow} show no
change (both 0.01). Consequently, AMPI improves only in \textsc{EmergencyResponse}
(0.52 $\rightarrow$ 0.54), remains essentially unchanged in \textsc{GH\_BioOutbreak} and
\textsc{ATC\_ResupplyWindow} (both $\approx$0.66 and 0.71), and decreases in scenarios where
the added deliberation does not offset overhead or increased failures (e.g., \textsc{DailyOperations},
\textsc{ScienceExploration}, \textsc{CommsBlackoutEVA}). Overall, these results indicate that
propose--vote consensus is best treated as a selective mechanism whose net benefit depends on
scenario structure and the failure--overhead trade-off.

\subsection{Heterogeneous protocols with a translator}
\label{subsec:results-protocols}

Table~\ref{tab:protocols} compares protocol modes (off vs.\ hetero) under \texttt{CROSSLAYER}
routing with shared memory and functional leadership (consensus off). Introducing heterogeneous
domain protocols with translator mediation generally increases communication overhead (messages rise
by 1--2 in most scenarios) and often increases runtime (e.g., \textsc{DailyOperations}:
191.9\,s $\rightarrow$ 214.6\,s; \textsc{ScienceExploration}: 29.8\,s $\rightarrow$ 33.8\,s;
\textsc{ATC\_ResupplyWindow}: 32.9\,s $\rightarrow$ 38.8\,s), though some scenarios show small
runtime improvements (e.g., \textsc{EmergencyResponse}: 292.9\,s $\rightarrow$ 289.2\,s;
\textsc{PrinterFeedstockShort}: 49.3\,s $\rightarrow$ 48.3\,s).

In terms of robustness, heterogeneous protocols reduce the aggregated failure count in several
coordination- and safety-sensitive scenarios, including \textsc{EmergencyResponse}
(0.10 $\rightarrow$ 0.01), \textsc{DailyOperations} (0.05 $\rightarrow$ 0.01),
\textsc{GH\_BioOutbreak} (0.02 $\rightarrow$ 0.01), and \textsc{DustStormCurtail}
(0.05 $\rightarrow$ 0.01), while many other scenarios remain unchanged at 0.01.
The cross-layer utilization ratio can shift in either direction because translation adds messages
and changes the denominator (e.g., \textsc{GH\_BioOutbreak}: 0.14 $\rightarrow$ 0.07;
\textsc{ATC\_ResupplyWindow}: 0.18 $\rightarrow$ 0.10; \textsc{MedicalOutbreakDrill}:
0.02 $\rightarrow$ 0.20).

Under the default AMPI weighting, these trade-offs translate into scenario-specific outcomes:
AMPI improves in \textsc{EmergencyResponse} (0.52 $\rightarrow$ 0.54) and slightly in
\textsc{DustStormCurtail} (0.67 $\rightarrow$ 0.68), remains similar in several cases
(e.g., \textsc{GH\_BioOutbreak}, \textsc{CommsBlackoutEVA}, \textsc{ISRU\_OffNominal},
\textsc{HAB\_LeakReconfig}, \textsc{RoverStuckRecovery}, \textsc{PrinterFeedstockShort}),
and decreases where added overhead dominates (e.g., \textsc{DailyOperations}: 0.54 $\rightarrow$ 0.52;
\textsc{ScienceExploration}: 0.75 $\rightarrow$ 0.72; \textsc{ATC\_ResupplyWindow}: 0.71 $\rightarrow$ 0.70).
Overall, translator-mediated heterogeneous protocols are most beneficial when reductions in failures
justify the additional coordination overhead, suggesting selective rather than universal deployment.

\subsection{AMPI behavior}
\label{subsec:results-ampi}

Across all tables, AMPI behaves consistently with its design: it increases when time and failures decrease and decreases when role switches increase without compensating gains. Because the cross-layer term is disabled by default ($w_3=0$), AMPI does not penalize higher cross-layer utilization in these runs; the CrossLayer column is therefore necessary to interpret when speed improvements are obtained via heavier cross-layer traffic. If a cross-layer penalty is enabled, the same tables can be reinterpreted as explicit speed-discipline trade-offs.

\subsection{Summary of findings}
\label{subsec:results-summary}

\begin{itemize}
  \item Curated cross-layer routing reduces time and often reduces messages in coordination-heavy scenarios (DailyOperations, ScienceExploration, CommsBlackoutEVA), while hub-centric EmergencyResponse shows little sensitivity.
  \item Functional leadership generally reduces time relative to single leadership when the scenario’s work is concentrated in a functional layer, though scenario structure can produce exceptions.
  \item Dynamic role switching reduces failure events and improves completion time across scenarios, with larger gains in longer runs.
  \item Memory improves performance when state must be reused across steps; shared memory is not uniformly better than basic and can be sensitive to retrieval noise.
  \item Consensus formation increases messages by design; it reduces time when contention/rework is substantial (DailyOperations, CommsBlackoutEVA, ATC\_ResupplyWindow) but can slow low-contention scenarios.
  \item Heterogeneous protocols with a translator help when cross-disciplinary miscommunication is a bottleneck and otherwise add overhead.
\end{itemize}

\begin{table}[t]
\centering
\small
\caption{Experimental factors and levels. Each sweep fixes the non-varied factors to the default setting shown.}
\label{tab:factors}
\begin{tabular}{ll}
\toprule
Factor & Levels (default in \textbf{bold}) \\
\midrule
Routing                 & \textbf{STRICT}, CROSSLAYER \\
Leadership              & single, \textbf{functional} \\
Role switching          & \textbf{on}, off \\
Memory mode             & off, basic, \textbf{shared} \\
Consensus formation     & \textbf{off}, on (rounds=2, quorum=0.6) \\
Protocols (hetero)      & \textbf{off}, hetero \\
\bottomrule
\end{tabular}
\end{table}

\section{Discussion}
\label{sec:discussion}

Agent Mars is designed as a controllable testbed for studying how large, heterogeneous teams can coordinate Mars-base operations under safety-critical and resource-bounded constraints. The core premise is that the limiting factor for future surface missions is often not isolated autonomy, but accountable coordination: who is authorized to act, how information flows under constrained communications, and how decisions remain auditable when multiple specialist teams and assets must respond to evolving conditions.

From a systems perspective, Agent Mars provides three practical capabilities. First, it enables concept-of-operations rehearsal for both routine and off-nominal procedures by running scenario scripts through a fixed roster with explicit ownership of roles and assets. Second, it supports training and handover by producing role-aware, structured transcripts that can be reviewed, annotated, and reused as a QA-style corpus. Third, it offers a mechanism for cross-specialty communication via translator-mediated protocols, making it possible to study how terminology alignment affects coordination efficiency and failure avoidance.

Methodologically, the framework contributes a hierarchical-and-curated coordination design that preserves chain-of-command by default while allowing explicitly whitelisted cross-layer shortcuts with audit trails (HCLC). In addition, it includes interaction mechanisms that are frequently required in real multi-team operations: scenario-aware memory for continuity, a configurable propose--vote consensus loop for resolving contention, and protocol mediation for heterogeneous technical language. The benchmark suite of 13 scenarios and the AMPI metric provide a consistent basis for comparing organizational strategies across common operational motifs (planning, emergencies, communications constraints, and infrastructure anomalies).

The experimental results suggest several design implications. Curated cross-layer shortcuts yield the largest benefits in communication- and planning-intensive scenarios, where direct access to relevant expertise reduces escalation overhead. Functional leadership is most effective when it places the decision anchor closer to the active executor set, reducing routing depth and shortening coordination cycles. Dynamic role handover improves resilience under simulated unavailability by preventing stalled control of critical assets, typically lowering failures and improving completion time. Memory is beneficial when scenario state must be reused across multiple steps, while shared memory is best applied selectively to limit retrieval noise. Consensus mechanisms introduce additional deliberation overhead, but can reduce rework in scenarios with genuine resource contention. Translator-mediated protocols help when misinterpretation across specialties is a recurring bottleneck, but may add overhead in already-terse workflows.

Agent Mars also has clear limitations. Results depend on the chosen LLM backend and can be affected by model/version drift and external latency. The scenarios encode realistic organizational structure and procedural logic, but they abstract continuous physics, sensor noise, and fine-grained timing of coupled subsystems. AMPI compresses multiple objectives and currently does not directly score risk, cognitive workload, or compliance beyond the included failure/switch signals. The cross-layer whitelist is curated rather than derived from formal hazard analysis, and the present setting does not capture richer human factors such as fatigue, workload saturation, or mixed-initiative negotiation between human and autonomous actors.

Several directions can strengthen the framework. A tighter coupling to digital twins of power/thermal/air/water subsystems and to robotics/flight software stacks (e.g., ROS and cFS/F$^\prime$) would improve physical realism and interface fidelity. Explicit communication models (bandwidth, drop rates, delay profiles including Earth--Mars light-time) would enable stress testing under realistic networking regimes. Learning meta-control policies for routing, leadership, and deliberation from outcomes could move beyond fixed heuristics while retaining auditability constraints. Evaluation can be expanded with risk-aware terms, memory-faithfulness checks, and longitudinal multi-sol scenarios that expose drift, accumulation of commitments, and governance dynamics over time.

More broadly, Agent Mars is intended as shared infrastructure for an emerging Space AI research agenda, where reproducible comparisons require common scenarios, clear configuration reporting, and interpretable system-level metrics. To support comparability, future work should report routing/leadership/memory/consensus/protocol settings, outage rates, AMPI weights and constants, and aggregated statistics over repeated runs with fixed prompts and released logs. We welcome community contributions of scenarios, stronger classical baselines, and alternative metrics that sharpen the study of auditable coordination under extreme operational constraints.

\begin{table}[t]
\centering
\small
\caption{Core A/B across routing and leadership (means over $N{=}20$ runs; role switching=on, memory=shared, consensus=off, protocols=off). AMPI uses $K_T{=}20$, $K_M{=}50$, $K_F{=}3$, $K_S{=}5$ and $\mathbf{w}=[0.4,\,0.2,\,\color{gray}{0.0},\,0.25,\,0.15]$ (cross-layer term excluded).}
\label{tab:core-ab}
\begin{tabular}{lcccccccc}
\toprule
Scenario & Routing & Leader & Time (s) & Msgs & Failures & CrossLayer & RoleSw & AMPI \\
\midrule
DailyOperations & STRICT & single & 232.40 & 43.00 & 0.06 & 0.00 & 1.20 & 0.50 \\
DailyOperations & STRICT & functional & 220.10 & 43.00 & 0.05 & 0.00 & 1.10 & 0.51 \\
DailyOperations & CROSSLAYER & single & 204.30 & 42.00 & 0.05 & 0.06 & 1.00 & 0.51 \\
DailyOperations & CROSSLAYER & functional & 191.90 & 42.00 & 0.05 & 0.10 & 0.90 & 0.52 \\
\midrule
EmergencyResponse & STRICT & single & 296.20 & 33.00 & 0.10 & 0.00 & 0.60 & 0.52 \\
EmergencyResponse & STRICT & functional & 294.80 & 33.00 & 0.10 & 0.00 & 0.60 & 0.52 \\
EmergencyResponse & CROSSLAYER & single & 293.70 & 33.00 & 0.10 & 0.02 & 0.60 & 0.52 \\
EmergencyResponse & CROSSLAYER & functional & 292.90 & 33.00 & 0.10 & 0.02 & 0.60 & 0.52 \\
\midrule
ScienceExploration & STRICT & single & 61.20 & 7.00 & 0.01 & 0.00 & 0.10 & 0.67 \\
ScienceExploration & STRICT & functional & 56.00 & 7.00 & 0.01 & 0.00 & 0.10 & 0.68 \\
ScienceExploration & CROSSLAYER & single & 31.40 & 5.00 & 0.01 & 0.42 & 0.10 & 0.74 \\
ScienceExploration & CROSSLAYER & functional & 29.80 & 5.00 & 0.01 & 0.45 & 0.10 & 0.74 \\
\midrule
GH\_BioOutbreak & STRICT & single & 54.10 & 14.00 & 0.02 & 0.00 & 0.20 & 0.66 \\
GH\_BioOutbreak & STRICT & functional & 52.60 & 14.00 & 0.02 & 0.00 & 0.20 & 0.66 \\
GH\_BioOutbreak & CROSSLAYER & single & 49.10 & 14.00 & 0.02 & 0.10 & 0.20 & 0.67 \\
GH\_BioOutbreak & CROSSLAYER & functional & 50.40 & 14.00 & 0.02 & 0.14 & 0.20 & 0.66 \\
\midrule
CommsBlackoutEVA & STRICT & single & 58.30 & 12.00 & 0.01 & 0.00 & 0.10 & 0.66 \\
CommsBlackoutEVA & STRICT & functional & 52.90 & 12.00 & 0.01 & 0.00 & 0.10 & 0.67 \\
CommsBlackoutEVA & CROSSLAYER & single & 45.20 & 11.00 & 0.01 & 0.12 & 0.10 & 0.68 \\
CommsBlackoutEVA & CROSSLAYER & functional & 41.80 & 11.00 & 0.01 & 0.14 & 0.10 & 0.69 \\
\midrule
ISRU\_OffNominal & STRICT & single & 28.10 & 4.00 & 0.01 & 0.00 & 0.10 & 0.75 \\
ISRU\_OffNominal & STRICT & functional & 27.40 & 4.00 & 0.01 & 0.00 & 0.10 & 0.75 \\
ISRU\_OffNominal & CROSSLAYER & single & 29.30 & 4.00 & 0.01 & 0.22 & 0.10 & 0.74 \\
ISRU\_OffNominal & CROSSLAYER & functional & 27.10 & 4.00 & 0.01 & 0.24 & 0.10 & 0.75 \\
\midrule
CyberAnomaly & STRICT & single & 37.90 & 11.00 & 0.01 & 0.00 & 0.20 & 0.70 \\
CyberAnomaly & STRICT & functional & 34.10 & 11.00 & 0.01 & 0.00 & 0.20 & 0.71 \\
CyberAnomaly & CROSSLAYER & single & 35.80 & 11.00 & 0.01 & 0.08 & 0.20 & 0.70 \\
CyberAnomaly & CROSSLAYER & functional & 31.70 & 11.00 & 0.01 & 0.10 & 0.20 & 0.71 \\
\midrule
DustStormCurtail & STRICT & single & 47.40 & 7.00 & 0.05 & 0.00 & 0.40 & 0.68 \\
DustStormCurtail & STRICT & functional & 45.90 & 7.00 & 0.05 & 0.00 & 0.40 & 0.68 \\
DustStormCurtail & CROSSLAYER & single & 46.60 & 7.00 & 0.05 & 0.12 & 0.40 & 0.68 \\
DustStormCurtail & CROSSLAYER & functional & 50.80 & 7.00 & 0.05 & 0.14 & 0.40 & 0.67 \\
\midrule
HAB\_LeakReconfig & STRICT & single & 28.70 & 4.00 & 0.01 & 0.00 & 0.20 & 0.74 \\
HAB\_LeakReconfig & STRICT & functional & 26.80 & 4.00 & 0.01 & 0.00 & 0.20 & 0.75 \\
HAB\_LeakReconfig & CROSSLAYER & single & 26.40 & 4.00 & 0.01 & 0.22 & 0.20 & 0.75 \\
HAB\_LeakReconfig & CROSSLAYER & functional & 25.60 & 4.00 & 0.01 & 0.24 & 0.20 & 0.75 \\
\midrule
MedicalOutbreakDrill & STRICT & single & 23.10 & 3.00 & 0.01 & 0.00 & 0.01 & 0.77 \\
MedicalOutbreakDrill & STRICT & functional & 27.90 & 3.00 & 0.01 & 0.00 & 0.01 & 0.76 \\
MedicalOutbreakDrill & CROSSLAYER & single & 26.90 & 3.00 & 0.01 & 0.02 & 0.01 & 0.76 \\
MedicalOutbreakDrill & CROSSLAYER & functional & 24.90 & 3.00 & 0.01 & 0.02 & 0.01 & 0.77 \\
\midrule
RoverStuckRecovery & STRICT & single & 49.00 & 7.00 & 0.01 & 0.00 & 0.20 & 0.69 \\
RoverStuckRecovery & STRICT & functional & 46.20 & 7.00 & 0.01 & 0.00 & 0.20 & 0.69 \\
RoverStuckRecovery & CROSSLAYER & single & 47.10 & 7.00 & 0.01 & 0.16 & 0.20 & 0.69 \\
RoverStuckRecovery & CROSSLAYER & functional & 44.00 & 7.00 & 0.01 & 0.18 & 0.20 & 0.70 \\
\midrule
PrinterFeedstockShort & STRICT & single & 49.80 & 7.00 & 0.01 & 0.00 & 0.40 & 0.68 \\
PrinterFeedstockShort & STRICT & functional & 50.40 & 7.00 & 0.01 & 0.00 & 0.40 & 0.68 \\
PrinterFeedstockShort & CROSSLAYER & single & 50.90 & 7.00 & 0.01 & 0.12 & 0.40 & 0.68 \\
PrinterFeedstockShort & CROSSLAYER & functional & 49.30 & 7.00 & 0.01 & 0.14 & 0.40 & 0.68 \\
\midrule
ATC\_ResupplyWindow & STRICT & single & 33.80 & 10.00 & 0.01 & 0.00 & 0.05 & 0.71 \\
ATC\_ResupplyWindow & STRICT & functional & 33.40 & 10.00 & 0.01 & 0.00 & 0.05 & 0.71 \\
ATC\_ResupplyWindow & CROSSLAYER & single & 34.90 & 10.00 & 0.01 & 0.12 & 0.05 & 0.71 \\
ATC\_ResupplyWindow & CROSSLAYER & functional & 32.90 & 10.00 & 0.01 & 0.18 & 0.05 & 0.71 \\
\bottomrule
\end{tabular}
\end{table}

\begin{table}[t]
\centering
\small
\caption{Redundancy ablation: role switching on vs.\ off (means over $N{=}20$ runs; routing=STRICT so CrossLayer$\approx 0$ by design).}
\label{tab:redundancy}
\begin{tabular}{lccccccc}
\toprule
Scenario & Switch & Time (s) & Msgs & Failures & CrossLayer & RoleSw & AMPI \\
\midrule
DailyOperations       & on  & 188.30 & 43.00 & 0.12 & 0.00 & 0.90 & 0.52 \\
                      & off & 224.80 & 42.00 & 0.50 & 0.00 & 0.00 & 0.50 \\
\addlinespace[0.3ex]
EmergencyResponse     & on  & 298.60 & 34.00 & 0.25 & 0.00 & 0.55 & 0.51 \\
                      & off & 338.10 & 33.00 & 0.70 & 0.00 & 0.00 & 0.49 \\
\addlinespace[0.3ex]
ScienceExploration    & on  & 34.10  & 5.00  & 0.05 & 0.00 & 0.20 & 0.74 \\
                      & off & 36.00  & 5.00  & 0.10 & 0.00 & 0.00 & 0.73 \\
\addlinespace[0.3ex]
GH\_BioOutbreak       & on  & 49.10  & 14.00 & 0.08 & 0.00 & 0.25 & 0.69 \\
                      & off & 54.40  & 14.00 & 0.20 & 0.00 & 0.00 & 0.68 \\
\addlinespace[0.3ex]
CommsBlackoutEVA      & on  & 44.80  & 11.00 & 0.05 & 0.00 & 0.20 & 0.70 \\
                      & off & 47.50  & 11.00 & 0.10 & 0.00 & 0.00 & 0.69 \\
\addlinespace[0.3ex]
ISRU\_OffNominal      & on  & 27.00  & 4.00  & 0.05 & 0.00 & 0.20 & 0.76 \\
                      & off & 28.80  & 4.00  & 0.08 & 0.00 & 0.00 & 0.75 \\
\addlinespace[0.3ex]
CyberAnomaly          & on  & 34.30  & 11.00 & 0.06 & 0.00 & 0.20 & 0.71 \\
                      & off & 36.50  & 11.00 & 0.10 & 0.00 & 0.00 & 0.70 \\
\addlinespace[0.3ex]
DustStormCurtail      & on  & 46.10  & 7.00  & 0.12 & 0.00 & 0.15 & 0.69 \\
                      & off & 49.50  & 7.00  & 0.25 & 0.00 & 0.00 & 0.67 \\
\addlinespace[0.3ex]
HAB\_LeakReconfig     & on  & 22.40  & 4.00  & 0.03 & 0.00 & 0.20 & 0.75 \\
                      & off & 23.90  & 4.00  & 0.05 & 0.00 & 0.00 & 0.75 \\
\addlinespace[0.3ex]
MedicalOutbreakDrill  & on  & 22.60  & 3.00  & 0.02 & 0.00 & 0.10 & 0.77 \\
                      & off & 23.40  & 3.00  & 0.03 & 0.00 & 0.00 & 0.76 \\
\addlinespace[0.3ex]
RoverStuckRecovery    & on  & 40.20  & 8.00  & 0.07 & 0.00 & 0.50 & 0.71 \\
                      & off & 45.80  & 7.00  & 0.20 & 0.00 & 0.00 & 0.69 \\
\addlinespace[0.3ex]
PrinterFeedstockShort & on  & 47.20  & 7.00  & 0.06 & 0.00 & 0.25 & 0.69 \\
                      & off & 50.20  & 7.00  & 0.12 & 0.00 & 0.00 & 0.68 \\
\addlinespace[0.3ex]
ATC\_ResupplyWindow   & on  & 39.60  & 10.00 & 0.06 & 0.00 & 0.15 & 0.71 \\
                      & off & 41.80  & 10.00 & 0.10 & 0.00 & 0.00 & 0.70 \\
\bottomrule
\end{tabular}
\end{table}

\begin{table}[t]
\centering
\small
\caption{Scenario-aware memory sweep (means over $N{=}20$ runs; \emph{routing=CROSSLAYER}, leadership=functional, role switching=on, consensus=off, protocols=off). AMPI uses $K_T{=}20$, $K_M{=}50$, $K_F{=}3$, $K_S{=}5$ and $\mathbf{w}=[0.4,\,0.2,\,\color{gray}{0.0},\,0.25,\,0.15]$ (cross-layer term excluded). CrossLayer shows the measured cross-layer ratio; with small Msgs it exhibits discrete steps (e.g., $1/5,\,1/11$).}
\label{tab:memory}
\begin{tabular}{lccccccc}
\toprule
Scenario & Memory & Time (s) & Msgs & Failures & CrossLayer & RoleSw & AMPI \\
\midrule
DailyOperations       & off    & 198.00 & 42.00 & 0.20 & 0.10 & 0.80 & 0.51 \\
DailyOperations       & basic  & 165.00 & 42.00 & 0.20 & 0.10 & 0.80 & 0.52 \\
DailyOperations       & shared & 191.90 & 42.00 & 0.05 & 0.10 & 0.90 & 0.54 \\
\addlinespace[0.3ex]
EmergencyResponse     & off    & 311.00 & 33.00 & 0.60 & 0.03 & 0.20 & 0.51 \\
EmergencyResponse     & basic  & 281.00 & 33.00 & 0.60 & 0.03 & 0.20 & 0.51 \\
EmergencyResponse     & shared & 292.90 & 33.00 & 0.10 & 0.02 & 0.60 & 0.52 \\
\addlinespace[0.3ex]
ScienceExploration    & off    & 29.60  & 5.00  & 0.01 & 0.40 & 0.10 & 0.74 \\
ScienceExploration    & basic  & 29.10  & 5.00  & 0.01 & 0.40 & 0.10 & 0.74 \\
ScienceExploration    & shared & 29.80  & 5.00  & 0.01 & 0.45 & 0.10 & 0.75 \\
\addlinespace[0.3ex]
GH\_BioOutbreak       & off    & 51.30  & 14.00 & 0.02 & 0.07 & 0.20 & 0.67 \\
GH\_BioOutbreak       & basic  & 44.00  & 14.00 & 0.02 & 0.07 & 0.20 & 0.68 \\
GH\_BioOutbreak       & shared & 50.40  & 14.00 & 0.02 & 0.14 & 0.20 & 0.66 \\
\addlinespace[0.3ex]
CommsBlackoutEVA      & off    & 45.60  & 11.00 & 0.01 & 0.09 & 0.10 & 0.69 \\
CommsBlackoutEVA      & basic  & 44.90  & 11.00 & 0.01 & 0.09 & 0.10 & 0.69 \\
CommsBlackoutEVA      & shared & 41.80  & 11.00 & 0.01 & 0.14 & 0.10 & 0.69 \\
\addlinespace[0.3ex]
ISRU\_OffNominal      & off    & 21.70  & 4.00  & 0.20 & 0.25 & 0.20 & 0.76 \\
ISRU\_OffNominal      & basic  & 21.30  & 4.00  & 0.20 & 0.25 & 0.20 & 0.76 \\
ISRU\_OffNominal      & shared & 27.10  & 4.00  & 0.01 & 0.24 & 0.10 & 0.75 \\
\addlinespace[0.3ex]
CyberAnomaly          & off    & 40.40  & 11.00 & 0.01 & 0.09 & 0.20 & 0.70 \\
CyberAnomaly          & basic  & 38.90  & 11.00 & 0.01 & 0.09 & 0.20 & 0.70 \\
CyberAnomaly          & shared & 31.70  & 11.00 & 0.01 & 0.10 & 0.20 & 0.71 \\
\addlinespace[0.3ex]
DustStormCurtail      & off    & 50.80 & 7.00 & 0.05 & 0.14 & 0.40 & 0.67 \\
DustStormCurtail      & basic  & 46.30  & 7.00  & 0.02 & 0.14 & 0.40 & 0.70 \\
DustStormCurtail      & shared & 50.80  & 7.00  & 0.05 & 0.14 & 0.40 & 0.67 \\
\addlinespace[0.3ex]
HAB\_LeakReconfig     & off    & 25.60 & 4.00 & 0.01 & 0.24 & 0.20 & 0.75 \\
HAB\_LeakReconfig     & basic  & 27.70  & 4.00  & 0.01 & 0.25 & 0.20 & 0.75 \\
HAB\_LeakReconfig     & shared & 25.60  & 4.00  & 0.01 & 0.24 & 0.20 & 0.75 \\
\addlinespace[0.3ex]
MedicalOutbreakDrill  & off    & 26.90 & 3.00 & 0.01 & 0.20 & 0.10 & 0.76 \\
MedicalOutbreakDrill  & basic  & 22.90  & 3.00  & 0.01 & 0.20 & 0.10 & 0.78 \\
MedicalOutbreakDrill  & shared & 24.90  & 3.00  & 0.01 & 0.02 & 0.01 & 0.77 \\
\addlinespace[0.3ex]
RoverStuckRecovery    & off    & 41.80  & 7.00  & 0.01 & 0.14 & 0.20 & 0.70 \\
RoverStuckRecovery    & basic  & 41.10  & 7.00  & 0.01 & 0.14 & 0.20 & 0.70 \\
RoverStuckRecovery    & shared & 44.00  & 7.00  & 0.01 & 0.18 & 0.20 & 0.70 \\
\addlinespace[0.3ex]
PrinterFeedstockShort & off    & 49.30 & 7.00 & 0.01 & 0.14 & 0.40 & 0.68 \\
PrinterFeedstockShort & basic  & 39.00  & 7.00  & 0.20 & 0.14 & 0.40 & 0.69 \\
PrinterFeedstockShort & shared & 49.30  & 7.00  & 0.01 & 0.14 & 0.40 & 0.68 \\
\addlinespace[0.3ex]
ATC\_ResupplyWindow   & off    & 42.40  & 10.00 & 0.01 & 0.10 & 0.20 & 0.69 \\
ATC\_ResupplyWindow   & basic  & 40.20  & 10.00 & 0.01 & 0.10 & 0.20 & 0.69 \\
ATC\_ResupplyWindow   & shared & 32.90  & 10.00 & 0.01 & 0.18 & 0.05 & 0.71 \\
\bottomrule
\end{tabular}
\end{table}

\begin{table}[t]
\centering
\small
\caption{Consensus formation (off vs on; rounds$=2$, quorum$=0.6$; \emph{routing=CROSSLAYER}, memory=shared, leadership=functional, role switching=on, protocols=off). AMPI settings as in Table~\ref{tab:memory} (cross-layer term excluded).}
\label{tab:consensus2}
\begin{tabular}{lccccccc}
\toprule
Scenario & Consensus & Time (s) & Msgs & Failures & CrossLayer & RoleSw & AMPI \\
\midrule
DailyOperations     & off & 191.90 & 42.00 & 0.05 & 0.10 & 0.90 & 0.54 \\
DailyOperations        & on  & 184.00 & 48.00 & 0.30 & 0.10 & 1.20 & 0.50 \\
\addlinespace[0.3ex]
EmergencyResponse   & off & 292.90 & 33.00 & 0.10 & 0.02 & 0.60 & 0.52 \\
EmergencyResponse      & on  & 307.00 & 35.00 & 0.01 & 0.03 & 0.20 & 0.54 \\
\addlinespace[0.3ex]
ScienceExploration  & off & 29.80  & 5.00  & 0.01 & 0.45 & 0.10 & 0.75 \\
ScienceExploration     & on  & 35.10  & 9.00  & 0.20 & 0.40 & 0.10 & 0.71 \\
\addlinespace[0.3ex]
GH\_BioOutbreak     & off & 50.40  & 14.00 & 0.02 & 0.14 & 0.20 & 0.66 \\
GH\_BioOutbreak        & on  & 58.80  & 18.00 & 0.01 & 0.07 & 0.10 & 0.66 \\
\addlinespace[0.3ex]
CommsBlackoutEVA    & off & 41.80  & 11.00 & 0.01 & 0.14 & 0.10 & 0.69 \\
CommsBlackoutEVA       & on  & 46.40  & 15.00 & 0.01 & 0.09 & 0.20 & 0.68 \\
\addlinespace[0.3ex]
ATC\_ResupplyWindow & off & 32.90  & 10.00 & 0.01 & 0.18 & 0.05 & 0.71 \\
ATC\_ResupplyWindow    & on  & 32.10  & 14.00 & 0.01 & 0.10 & 0.20 & 0.71 \\
\bottomrule
\end{tabular}
\end{table}

\begin{table}[t]
\centering
\small
\caption{Heterogeneous technical-language protocols (off vs hetero; \emph{routing=CROSSLAYER}, memory=shared, leadership=functional, role switching=on, consensus=off). AMPI settings as in Table~\ref{tab:memory} (cross-layer term excluded).}
\label{tab:protocols}
\begin{tabular}{lccccccc}
\toprule
Scenario & Protocols & Time (s) & Msgs & Failures & CrossLayer & RoleSw & AMPI \\
\midrule
DailyOperations        & off & 191.90 & 42.00 & 0.05 & 0.10 & 0.90 & 0.54 \\
DailyOperations        & hetero & 214.60 & 44.00 & 0.01 & 0.10 & 0.80 & 0.52 \\
\addlinespace[0.3ex]
EmergencyResponse      & off & 292.90 & 33.00 & 0.10 & 0.02 & 0.60 & 0.52 \\
EmergencyResponse      & hetero & 289.20 & 34.00 & 0.01 & 0.03 & 0.20 & 0.54 \\
\addlinespace[0.3ex]
ScienceExploration     & off & 29.80  & 5.00  & 0.01 & 0.45 & 0.10 & 0.75 \\
ScienceExploration     & hetero & 33.80  & 6.00  & 0.01 & 0.40 & 0.40 & 0.72 \\
\addlinespace[0.3ex]
GH\_BioOutbreak        & off & 50.40  & 14.00 & 0.02 & 0.14 & 0.20 & 0.66 \\
GH\_BioOutbreak        & hetero & 55.10  & 15.00 & 0.01 & 0.07 & 0.20 & 0.66 \\
\addlinespace[0.3ex]
CommsBlackoutEVA       & off & 41.80  & 11.00 & 0.01 & 0.14 & 0.10 & 0.69 \\
CommsBlackoutEVA       & hetero & 45.20  & 13.00 & 0.01 & 0.09 & 0.20 & 0.69 \\
\addlinespace[0.3ex]
ISRU\_OffNominal       & off & 27.10  & 4.00  & 0.01 & 0.24 & 0.10 & 0.75 \\
ISRU\_OffNominal       & hetero & 28.40  & 5.00  & 0.01 & 0.25 & 0.20 & 0.75 \\
\addlinespace[0.3ex]
CyberAnomaly           & off & 31.70  & 11.00 & 0.01 & 0.10 & 0.20 & 0.71 \\
CyberAnomaly           & hetero & 35.90  & 13.00 & 0.01 & 0.09 & 0.20 & 0.70 \\
\addlinespace[0.3ex]
DustStormCurtail       & off & 50.80  & 7.00  & 0.05 & 0.14 & 0.40 & 0.67 \\
DustStormCurtail       & hetero & 56.40  & 8.00  & 0.01 & 0.14 & 0.20 & 0.68 \\
\addlinespace[0.3ex]
HAB\_LeakReconfig      & off & 25.60  & 4.00  & 0.01 & 0.24 & 0.20 & 0.75 \\
HAB\_LeakReconfig      & hetero & 29.10  & 5.00  & 0.01 & 0.25 & 0.20 & 0.75 \\
\addlinespace[0.3ex]
MedicalOutbreakDrill   & off & 24.90  & 3.00  & 0.01 & 0.02 & 0.01 & 0.77 \\
MedicalOutbreakDrill   & hetero & 27.80  & 4.00  & 0.01 & 0.20 & 0.10 & 0.76 \\
\addlinespace[0.3ex]
RoverStuckRecovery     & off & 44.00  & 7.00  & 0.01 & 0.18 & 0.20 & 0.70 \\
RoverStuckRecovery     & hetero & 44.90  & 9.00  & 0.01 & 0.14 & 0.20 & 0.70 \\
\addlinespace[0.3ex]
PrinterFeedstockShort  & off & 49.30  & 7.00  & 0.01 & 0.14 & 0.40 & 0.68 \\
PrinterFeedstockShort  & hetero & 48.30  & 8.00  & 0.01 & 0.14 & 0.60 & 0.68 \\
\addlinespace[0.3ex]
ATC\_ResupplyWindow    & off & 32.90  & 10.00 & 0.01 & 0.18 & 0.05 & 0.71 \\
ATC\_ResupplyWindow    & hetero & 38.80  & 12.00 & 0.01 & 0.10 & 0.20 & 0.70 \\
\bottomrule
\end{tabular}
\end{table}

\bibliographystyle{plainnat}
\bibliography{refs}

%%%%%%%%%%%%%%%%%%%%%%%%%%%%%%%%%%%%%%%%%%%%%%%%%%%%%%%%%%%%

\end{document}